\documentclass[twoside,twocolumn,9pt]{article}
\usepackage{extsizes}
\usepackage[super,sort&compress,comma]{natbib} 
\usepackage[version=3]{mhchem}
\usepackage[left=1.5cm, right=1.5cm, top=1.785cm, bottom=2.0cm]{geometry}
\usepackage{balance}
\usepackage{times,mathptmx}
\usepackage{sectsty}
\usepackage{graphicx} 
\usepackage{lastpage}
\usepackage[format=plain,justification=justified,singlelinecheck=false,font={stretch=1.125,small,sf},labelfont=bf,labelsep=space]{caption}
\usepackage{float}
\usepackage{fancyhdr}
\usepackage{fnpos}
\usepackage[USenglish]{babel}
\usepackage{array}
\usepackage{droidsans}
\usepackage{charter}
\usepackage[T1]{fontenc}
\usepackage[usenames,dvipsnames]{xcolor}
\usepackage{setspace}
\usepackage[compact]{titlesec}
\usepackage[pdfusetitle]{hyperref}

\usepackage{epstopdf}

\definecolor{cream}{RGB}{222,217,201}

\usepackage{amsmath}
\usepackage{amssymb}
\allowdisplaybreaks
\usepackage{bbold}
\usepackage{ragged2e}
\usepackage{widetext}

\IfFileExists{.git/modules/bibtex/hooks/commit-msg.sample}{
    \usepackage{tikz}
    \usepackage{pgfplots}
    \usepackage{gnuplot-lua-tikz}
    \usetikzlibrary{external}
    \usepgfplotslibrary{external}
    \tikzexternalize[prefix=tikz/]

    \usepackage{todonotes}
}{
    \usepackage{tikzexternal}
    \tikzexternalize
    \IfFileExists{tikz/\jobname-figure0.pdf}{
        \tikzsetexternalprefix{tikz/}
    }{
        \relax
    }
    \let\tikzexternaldisable\relax
    \let\tikzexternalenable\relax

    \newcommand{\todo}[2][]{}
}
\def\gpsetlinewidth#1{\ifdim#1 pt>0pt \pgfsetlinewidth{#1\gpbaselw}\fi}

\makeatletter
\def\input@path{{./}{plots/}}
\makeatother
\graphicspath{{./}{plots/}}

\usepackage{jabbrv}
\DefineSpuriousJournalWord{The}
\DefineSpuriousJournalWord{on}
\DefineSpuriousJournalWord{its}
\DefineJournalException{Chemistry: A European Journal}{Chem. Eur. J.}
\DefineJournalException{Journal of Physics: Condensed Matter}{J. Phys. Condens. Matter}
\DefineJournalException{Angewandte Chemie}{Angew. Chem.}
\DefineJournalException{Computers \& Fluids}{Comput. Fluids}
\DefineJournalException{Annalen der Physik}{Ann. Phys.}
\DefineJournalException{Arkiv f\"or matematik, astronomi och fysik}{Ark. Mat. Astron. Fys.}
\DefineJournalException{Physica A: Statistical Mechanics and its Applications}{Phys. A Stat. Mech. Appl.}
\DefineJournalAbbreviation{Surfaces}{Surf}
\DefineJournalAbbreviation{Reports}{Rep}
\DefineJournalAbbreviation{Physique}{Phys}
\DefineJournalAbbreviation{Transactions}{Trans}

\makeatletter
\@ifpackageloaded{tikz}{
    \renewcommand{\todo}[2][]{\tikzexternaldisable\@todo[#1]{#2}\tikzexternalenable}
}{
    \relax
}
\usepackage{xpatch}
\patchcmd{\f@nch@vbox}{\global}{}{}{}
\makeatother

\renewcommand{\vec}[1]{\mathbf{#1}}
\newcommand{\unitvec}[1]{\hat{\mathbf{#1}}}

\newcommand{\tgrid}{\tau}
\newcommand{\dr}{s}
\newcommand{\drv}{\vec{s}}

\usepackage{cleveref}
\usepackage{titling}

\title{Hydrodynamic Mobility Reversal of Squirmers near Flat and Curved Surfaces}

\date{\today}

\begin{document}

\pagestyle{fancy}
\thispagestyle{plain}
\fancypagestyle{plain}{

\fancyhead[C]{\includegraphics[width=18.5cm]{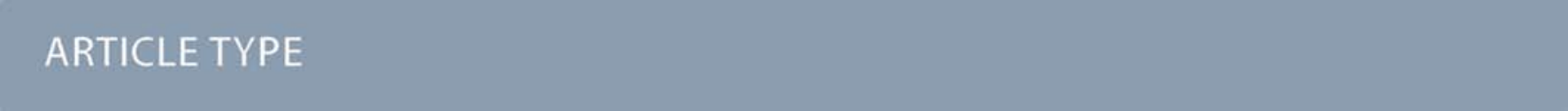}}
\fancyhead[L]{\hspace{0cm}\vspace{1.5cm}\includegraphics[height=30pt]{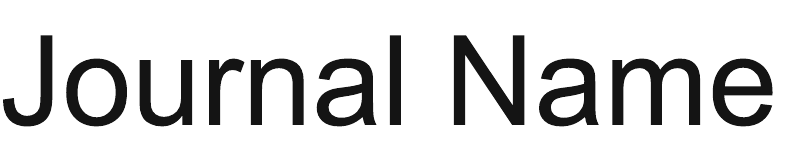}}
\fancyhead[R]{\hspace{0cm}\vspace{1.7cm}\includegraphics[height=55pt]{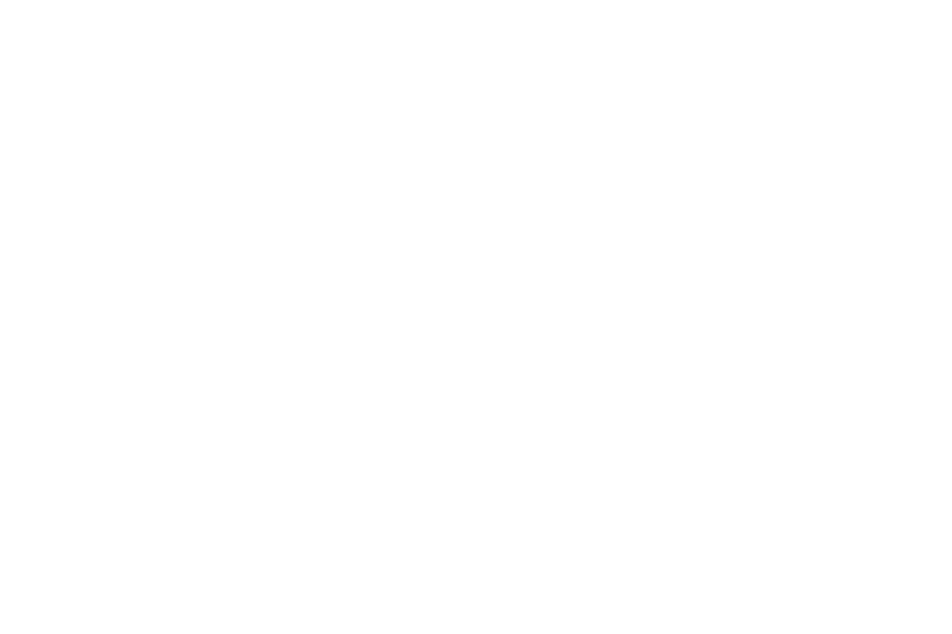}}
\renewcommand{\headrulewidth}{0pt}
}

\makeFNbottom
\makeatletter
\renewcommand\LARGE{\@setfontsize\LARGE{15pt}{17}}
\renewcommand\Large{\@setfontsize\Large{12pt}{14}}
\renewcommand\large{\@setfontsize\large{10pt}{12}}
\renewcommand\footnotesize{\@setfontsize\footnotesize{7pt}{10}}
\makeatother

\renewcommand{\thefootnote}{\fnsymbol{footnote}}
\renewcommand\footnoterule{\vspace*{1pt}%
\color{cream}\hrule width 3.5in height 0.4pt \color{black}\vspace*{5pt}} 
\setcounter{secnumdepth}{5}

\makeatletter 
\renewcommand\@biblabel[1]{#1}            
\renewcommand\@makefntext[1]%
{\noindent\makebox[0pt][r]{\@thefnmark\,}#1}
\makeatother 
\renewcommand{\figurename}{\small{Fig.}~}
\sectionfont{\sffamily\Large}
\subsectionfont{\normalsize}
\subsubsectionfont{\bf}
\setstretch{1.125} 
\setlength{\skip\footins}{0.8cm}
\setlength{\footnotesep}{0.25cm}
\setlength{\jot}{10pt}
\titlespacing*{\section}{0pt}{4pt}{4pt}
\titlespacing*{\subsection}{0pt}{15pt}{1pt}

\fancyfoot{}
\fancyfoot[LO,RE]{\vspace{-7.1pt}\includegraphics[height=9pt]{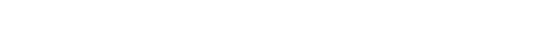}}
\fancyfoot[CO]{\vspace{-7.1pt}\hspace{13.2cm}\includegraphics{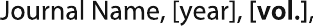}}
\fancyfoot[CE]{\vspace{-7.2pt}\hspace{-14.2cm}\includegraphics{head_foot/RF}}
\fancyfoot[RO]{\footnotesize{\sffamily{1--\pageref{LastPage} ~\textbar  \hspace{2pt}\thepage}}}
\fancyfoot[LE]{\footnotesize{\sffamily{\thepage~\textbar\hspace{3.45cm} 1--\pageref{LastPage}}}}
\fancyhead{}
\renewcommand{\headrulewidth}{0pt} 
\renewcommand{\footrulewidth}{0pt}
\setlength{\arrayrulewidth}{1pt}
\setlength{\columnsep}{6.5mm}
\setlength\bibsep{1pt}

\makeatletter 
\newlength{\figrulesep} 
\setlength{\figrulesep}{0.5\textfloatsep} 

\newcommand{\topfigrule}{\vspace*{-1pt}%
\noindent{\color{cream}\rule[-\figrulesep]{\columnwidth}{1.5pt}} }

\newcommand{\botfigrule}{\vspace*{-2pt}%
\noindent{\color{cream}\rule[\figrulesep]{\columnwidth}{1.5pt}} }

\newcommand{\dblfigrule}{\vspace*{-1pt}%
\noindent{\color{cream}\rule[-\figrulesep]{\textwidth}{1.5pt}} }

\makeatother

\twocolumn[
  \begin{@twocolumnfalse}
\vspace{3cm}
\sffamily
\begin{tabular}{m{4.5cm} p{13.5cm} }

\includegraphics{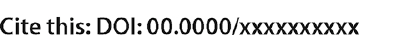} & \noindent\LARGE{\textbf{\thetitle}} \\
\vspace{0.3cm} & \vspace{0.3cm} \\

 & \noindent\large{Michael Kuron\textit{$^{\ast a \ddag}$}, Philipp St\"ark\textit{$^{a \ddag}$}, Christian Holm\textit{$^{a \ddag}$},  and Joost de Graaf\textit{$^{b \ddag}$}} \\

\includegraphics{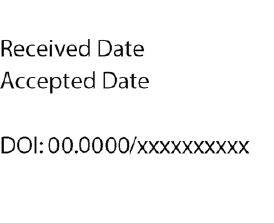} & \noindent\normalsize{%
Self-propelled particles have been experimentally shown to orbit spherical obstacles and move along surfaces.
Here, we theoretically and numerically investigate this behavior for a hydrodynamic squirmer interacting with spherical objects and flat walls using three different methods of approximately solving the Stokes equations:
The method of reflections, which is accurate in the far field; lubrication theory, which describes the close-to-contact behavior; and a lattice Boltzmann solver that accurately accounts for near-field flows.
The method of reflections predicts three distinct behaviors:
orbiting/sliding, scattering, and hovering, with orbiting being favored for lower curvature as in the literature.
Surprisingly, it also shows backward orbiting/sliding for sufficiently strong pushers, caused by fluid recirculation in the gap between the squirmer and the obstacle leading to strong forces opposing forward motion.
Lubrication theory instead suggests that only hovering is a stable point for the dynamics.
We therefore employ lattice Boltzmann to resolve this discrepancy and we qualitatively reproduce the richer far-field predictions.
Our results thus provide insight into a possible mechanism of mobility reversal mediated solely through hydrodynamic interactions with a surface.
} \\

\end{tabular}

 \end{@twocolumnfalse} \vspace{0.6cm}

  ]

\renewcommand*\rmdefault{bch}\normalfont\upshape
\rmfamily
\section*{}
\vspace{-1cm}


\footnotetext{\textit{$^{a}$~Institute for Computational Physics, University of Stuttgart, Allmandring 3, 70569 Stuttgart, Germany. E-mail: \texttt{mkuron@icp.uni-stuttgart.de}}}
\footnotetext{\textit{$^{b}$~Institute for Theoretical Physics, Center for Extreme Matter and Emergent Phenomena, Utrecht University, Princetonplein 5, 3584 CC Utrecht, The Netherlands.}}

\footnotetext{$\ddag$ 
Conceptualization: MK, JdG;
Calculations and Analysis: MK, PS, JdG;
Writing: MK, JdG;
Supervision: CH, JdG;
Funding Acquisition: CH, JdG;
Resources: CH.
}




\section{Introduction}
\label{sec:intro}

An increasing body of experimental and theoretical work demonstrates
that the proximity of surfaces has an important effect on the behavior
of self-propelled particles.  In biology,
spermatozoa\cite{woolley03a,friedrich10a,bukatin15a} and
bacteria\cite{berg90a} can circle near a flat wall, which has been
attributed to hydrodynamic interactions and the specifics of flagellar
beating\cite{lauga06a,lauga09a,elgeti10a,gadelha10a,giacche10a,alvarez14a,bechinger16a,saggiorato17a}.
Artificial self-propelled particles, which can move through the
catalytic decomposition of \ce{H2O2}\cite{paxton04a,howse07a}, i.e.,
chemical swimmers, also respond to the presence of a
surface\cite{volpe11a,takagi14a,das15a,brown16a,simmchen16a}.  In this
case, there can be both a
hydrodynamic\cite{felderhof77b,lauga06a,llopis10a,spagnolie12a,spagnolie15a}
and a chemical coupling to the
surface\cite{popescu09a,das15a,uspal15a,mozaffari16a,simmchen16a},
which themselves are intimately linked through the way they lead to
self-propulsion\cite{anderson89a,paxton04a,golestanian05a,wang06g,golestanian07a,howse07a,popescu10a,moran11a,ebbens12a,brown14a,brown17a,campbell18a-pre}.

Experimentally, chemical swimmers are well known to be orientationally
locked near a flat surface\cite{das15a,simmchen16a}.  This locking has
been linked to specifics of the reaction mechanism and the local
hydrodynamic interactions that it induces\cite{uspal15a,das15a}.
Chemical swimmers may also follow the surface topology.  For example,
they interact with small variations of the substrate's
height\cite{volpe11a,simmchen16a}, as has been qualitatively described
using simple theoretical model swimmers\cite{simmchen16a}.  In
addition, chemical patterning of the surface has been shown to
significantly modify the mobility of a chemical
swimmer\cite{hu15a,uspal16a,ceylan17a,popescu17a,uspal18a}.  These
man-made swimmers can also follow strongly curved surfaces, even
leading them to orbit around spherical
obstacles\cite{takagi14a,brown16a}.

The orbiting of swimmers has been studied extensively using
hydrodynamic descriptions\cite{spagnolie15a,chamolly17a}.  In the far
field, the associated hydrodynamic problem is typically solved using
the method-of-reflections approximation\cite{felderhof77b} and
Fax\'en's law\cite{faxen22b,brenner64a}.  \citet{spagnolie15a} account
for the leading-order hydrodynamic force-dipole moment in their
analysis and find that there is a critical radius for orbiting.  Only
pusher swimmers --- ones that have an extensile flow field --- enter
such a trajectory\cite{spagnolie15a}; pullers on the other hand are
trapped in a `hovering' state, wherein they point straight into the
surface.  However, the methods of reflections is known to break down
for small swimmer-obstacle separations\cite{spagnolie12a}.

In the lubrication regime, which captures the behavior for vanishing
gap sizes, a swimmer's ability to follow a path along a planar wall has been
examined\cite{lintuvuori16a,shen18a}.  Specifically, \citet{lintuvuori16a}
studied a squirmer, which is a simple model swimmer that accounts for
finite-size contributions to the flow field.  The results for a
squirmer near a flat wall may be readily transferred to orbiting around
objects with low curvature.  Unfortunately, lubrication theory does
not provide substantial insight other than for the hovering state,
wherein the swimmer's direction of motion is into the obstacle and no
tangential displacement occurs.

A combination of lubrication and far-field results is often used in an
attempt to bridge the gap between these two
regimes\cite{lintuvuori16a,chamolly17a,shen18a}.  This approximation gives
rise to steady orbiting pullers and oscillatory orbits for
pushers\cite{lintuvuori16a}, the latter of which are a result of a
competition between the two regimes.  The critical radius for orbiting
will also be reduced by this interplay\cite{chamolly17a}.

In the case of a flat wall, the intermediate regime has been resolved
using the boundary element (BEM) method\cite{ishimoto13a}, as well as
the lattice Boltzmann (LB) method\cite{lintuvuori16a,shen18a}, and
multi-particle collision dynamics\cite{schaar15a} (MPCD).
\citet{ishimoto13a} observe that a puller squirmer moves stably along
the wall, pointing slightly toward it.  \citet{lintuvuori16a}
reproduce the behaviors found in their analytic predictions, which
combine the far field and lubrication regimes.  However, this level of
analysis has not yet been performed for orbiting.

In this paper, we examine in-depth the effect of the surface curvature
on the hydrodynamic orbiting of a squirmer
and take the limit to the behavior near a flat surface.  To
efficiently explore parameter space, we employ far-field
approximations, as well as the LB
method\cite{mcnamara88a,higuera89a,ladd94a,krueger17a} that accurately
resolves the near-field flows. Using this approach we reproduce the three
behaviors reported in literature: orbiting around the sphere/sliding
along the wall, scattering, and
hovering\cite{ishimoto13a,spagnolie15a,lintuvuori16a,chamolly17a}.
Surprisingly, we find a second type of orbiting in both our
hydrodynamic approaches where the swimmer effectively moves in the
direction opposite to its bulk motion.  Backward orbiting appears
for strong pusher squirmers and supersedes the forward orbiting
predicted for a point-like dipole swimmer using identical hydrodynamic
parameters\cite{spagnolie15a}.  This behavior can be shown to result
from fluid recirculation in the gap between the squirmer and the
surface leading to strong forces opposing forward motion.
In the limit of a flat wall, we similarly find backward sliding.

The behavior of a squirmer near an obstacle is controlled by the
strength and sign of its dipole moment and is also sensitive to the
curvature of the obstacle.  Scattering takes place for sufficiently
neutral squirmers, which only have a source dipole flow field in bulk,
while hovering and orbiting require the presence of a force-dipole
contribution to the flow field.  We find that the minimal force-dipole
moment that leads to orbiting scales quadratically with the curvature
for puller squirmers.  Backward orbiting/sliding and hovering can be
suppressed by introducing short-ranged repulsions between the squirmer
and the surface.  However, backward motion supplants forward motion
even for imposed gap sizes of one tenth of the squirmer radius.

We will focus on hydrodynamic interactions here, but our results are
also of interest to the ongoing study of motion of chemical swimmers
near surfaces.  We will show that it is possible to reverse the
mobility of a swimmer by modifying its hydrodynamic force-dipole
moment without changing the bulk swim velocity.  This strong response
to the presence of a surface underpins the need for more experiments
performed in bulk in order to isolate the effect of environmental
changes on swimmer mobility.  Our predictions provide a stepping
stone toward understanding the richer behaviors encountered when
introducing coupling between solute gradients and hydrodynamic flow
fields\cite{michelin14a,ibrahim16a,popescu18a}.

The remainder of this paper is laid out as follows: \Cref{sec:theory}
introduces the squirmer model and describes the problem considered.
We also explain how the hydrodynamics are solved to obtain
trajectories of the squirmers.  In \cref{sec:characterization}, we
introduce a characterization of these trajectories, before we present
our main results in \cref{sec:results}, wherein we discuss the
influence of obstacle size, short-range repulsion, and higher-order
hydrodynamic moments.  We conclude and present an outlook in
\cref{sec:conclusion}.

\section{Model and Method}
\label{sec:theory}

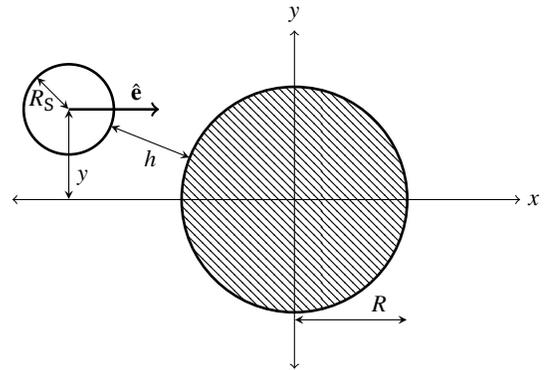
\begin{figure}[htb]
\centering
\begin{tikzpicture}
	\usetikzlibrary{calc}
	\usetikzlibrary{patterns}

\tikzset{guide/.style={densely dotted,shorten <=1pt,shorten >=1pt}}
\tikzset{vector/.style={->,line width=1pt}}
\tikzset{object/.style={line width=1pt}}
\tikzset{obstacle/.style={pattern=north west lines}}
\tikzset{length/.style={<->,>=stealth,shorten <=0.2pt,shorten >=0.2pt}}
\tikzset{axes/.style={<->}}

\pgfmathsetmacro\Rs{0.6}
\pgfmathsetmacro\R{1.5}
\pgfmathsetmacro\xoff{-2*\R}
\pgfmathsetmacro\yoff{0.8*\R}
\pgfmathsetmacro\Rmark{-60}
\pgfmathsetmacro\Rsmark{135}

\draw [axes] (-2.5*\R,0) -- (2*\R,0) node [right] {$x$};
\draw [axes] (0,-1.5*\R) -- (0,1.5*\R) node [above] {$y$};

\draw[object,obstacle] (0,0) circle(\R) {};
\draw [length] (0,-\R-0.1) -- node [above,pos=0.75] {$R$} ++(\R,0);

\draw [vector] (\xoff,\yoff) -- node[pos=0.75,above] {$\unitvec{e}$} ++(2*\Rs,0);

\draw[object] (\xoff,\yoff) circle(\Rs) {};
\draw [length] (\xoff,\yoff) -- node [right,pos=0.5*(\yoff+\Rs)/\yoff)] {$y$} (\xoff,0);
\coordinate (Rs) at ({\xoff+\Rs*cos(\Rsmark)},{\yoff+\Rs*sin(\Rsmark)});
\draw [length] (\xoff,\yoff) -- node [below left=-6pt] {$R_\text{S}$} (Rs);

\pgfmathsetmacro\ang{atan2(\yoff,\xoff)}
\coordinate (s) at ({\xoff+cos(\ang+180)*\Rs},{\yoff+sin(\ang+180)*\Rs});
\coordinate (o) at ({cos(\ang)*\R},{sin(\ang)*\R});
\draw [length] (s) -- node[below] {$h$} (o) {};

\end{tikzpicture}
\caption{ The geometry of the system investigated:  The obstacle of
  radius $R$ is located at the origin, while the squirmer of radius
  $R_\text{S}$ and orientation $\unitvec{e}$ is at $(x,y)^\intercal$.  The
  size of the gap between the two objects is $h$.  For the case of the
  flat wall ($R\rightarrow\infty)$, $y$ loses its meaning, but the
  other quantities remain well-defined.  }
\label{fig:geometry}
\end{figure}

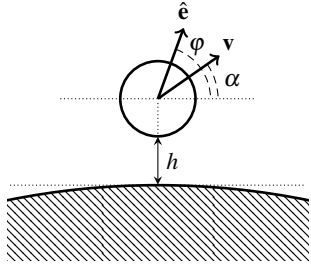
\begin{figure}[htb]
\centering
\begin{tikzpicture}
\usetikzlibrary{calc}
\usetikzlibrary{patterns}

\tikzset{guide/.style={densely dotted,shorten <=1pt,shorten >=1pt}}
\tikzset{vector/.style={->,line width=1pt}}
\tikzset{object/.style={line width=1pt}}
\tikzset{obstacle/.style={pattern=north west lines}}
\tikzset{length/.style={<->,>=stealth,shorten <=0.2pt,shorten >=0.2pt}}
\tikzset{angle/.style={densely dashed}}

\pgfmathsetmacro\Rs{0.5}
\pgfmathsetmacro\R{20*\Rs}
\pgfmathsetmacro\L{4*\Rs}
\pgfmathsetmacro\yoff{2.3*\Rs}
\pgfmathsetmacro\eang{70}
\pgfmathsetmacro\vang{35}
\pgfmathsetmacro\erad{0.2}
\pgfmathsetmacro\vrad{0.3}

\begin{scope}
	\clip (-\L,1) rectangle (\L,-1);
	\coordinate (o) at (0,-\R);
	\draw[object,obstacle] (o) circle(\R) {};
\end{scope}

\draw[guide] (-\L,0) -- (\L,0);

\coordinate (c) at (0,\yoff);
\draw[object] (c) circle(\Rs) {};

\draw[guide] (-\L/1.5,\yoff) -- (\L/1.5,\yoff);

\coordinate (e) at ({cos(\eang)*\Rs},{\yoff+sin(\eang)*\Rs});
\draw [vector] (0,\yoff) -- node[pos=1,above] {$\unitvec{e}$} ++(\eang:2*\Rs);
\draw [angle] (e) ++(\eang:\erad) arc(\eang:0:\Rs+\erad) node[pos=0.4,above] {$\varphi$};

\coordinate (v) at ({cos(\vang)*\Rs},{\yoff+sin(\vang)*\Rs});
\draw [vector] (0,\yoff) -- node[pos=0.9,above right] {$\vec{v}$} ++(\vang:2*\Rs);
\draw [angle] (v) ++(\vang:\vrad) arc(\vang:0:\Rs+\vrad) node[pos=0.5,right] {$\alpha$};

\draw [length] (0,0) -- node[right] {$h$} (0,\yoff-\Rs) {};
\draw [guide] (0,\yoff-\Rs) -- (0,\yoff) {};

\end{tikzpicture}
\caption{ The angles $\varphi$ and $\alpha$ are between the tangent
  plane at the closest point on the obstacle's surface and the
  squirmer's orientation $\unitvec{e}$ and direction of motion
  $\vec{v}$, respectively.  The length $h$ is the size of the gap
  between squirmer and obstacle.  }
\label{fig:angles}
\end{figure}

In this paper, we study the interaction of a squirmer of radius
$R_\text{S}$ with a spherical obstacle of radius $R$ (or a flat wall,
corresponding to $R\rightarrow\infty$) as illustrated in
\cref{fig:geometry}.  The squirmer is free to move and its position is
$\vec{r}_\text{S}=(x,y)^\intercal$ with the superscript denoting
transposition, while the obstacle is fixed at the origin.
\Cref{fig:angles} introduces the angles and distances used to describe
the squirmer's position and orientation: $\varphi$ is the angle
between the squirmer's orientation $\unitvec{e}$ and the tangent plane
at the closest point on the obstacle's surface; $\alpha$ is the angle
between the direction of the squirmer's velocity $\vec{v}$ and the
tangent plane; and $h$ is the size of the gap between the squirmer's
and the obstacle's surfaces.  Initially, the squirmer is located far
away from the obstacle --- we effectively take the limit to infinity
--- at different distances $y_0$ from the $x$-axis.  $\unitvec{e}$
points along the $x$-axis, corresponding to
$\varphi_0=\arcsin(y_0/R_\text{S})-90^\circ$.  For the case of the
flat wall, the squirmer starts at a large distance above the wall and
oriented at different $\varphi_0$ against it.  Subscript zeros ($_0$)
here refer to the respective variable at time $t=0$.

In the rest of this section, we describe a method to determine how the
model's parameters --- like dipolarity $\beta$, angle of incidence
$\varphi_0$, and size ratio $R/R_\text{S}$ --- affect the squirmer's
ability to enter into an orbit around or be scattered at the obstacle:
we \hyperref[subsec:squirmer]{(1)} introduce the squirmer model,
\hyperref[subsec:modes]{(2)} perform a hydrodynamic multipole
decomposition, \hyperref[subsec:reflect]{(3)} introduce the method of
reflections to account for the obstacle's presence,
\hyperref[subsec:faxen]{(4)} use Fax\'en's law to calculate the
response of the squirmer to fluid flow, and
\hyperref[subsec:numerical]{(5)} explain how to numerically integrate
the trajectory.  Finally, we \hyperref[subsec:lubrication]{(6)} use
lubrication theory to briefly discuss the stability of bound states
and \hyperref[subsec:LB]{(7)} introduce a numerical method that is
capable of resolving the near-field flow.

\subsection{Squirmers and the Stokes equation}
\label{subsec:squirmer}

Theoretical descriptions typically use simple swimmer models that
describe only the resulting hydrodynamic flow in a far-field
approximation, which eliminates the complex details of a
microswimmer's propulsion method.  As self-propulsion is
force-free\cite{ishikawa09a,lauga09a}, the lowest nonzero term of a
hydrodynamic multipole expansion is the dipole.  The force dipole
decays with distance as $r^{-2}$, which justifies the often-applied
truncation beyond this order.  Point dipoles are, however, difficult
to handle numerically, due to their inherent
divergences\cite{mathijssen15a,degraaf17a}, while extended dipoles are
inconvenient to analytical theory.

A widely used model that can account for finite sizes of the swimmer
is the squirmer model.  \citet{lighthill52a} originally introduced it
to explain swimming by an oscillatory shape change.  \citet{blake71b}
later used it to describe the microorganism \emph{Paramecium}, which
propels via a specific beat pattern of the cilia on its surface.  Both
authors expanded the flow at the swimmer's surface into spherical
harmonics and discovered that the first two modes are sufficient to
describe the resulting flow.  If the squirmer is impermeable, radial
flow through the surface can further be ignored\cite{ishikawa06a}, so
that the motion of cilia on the surface of a sphere of radius
$R_\text{S}$ can be described by the envelope\cite{blake71b}
\begin{equation}
\left.\vec{u}(\vec{r})\right|_{\dr=R_\text{S}}
=\left(B_1+B_2\frac{\unitvec{e}\cdot\drv}{\dr}\right)
\left(\frac{\unitvec{e}\cdot\drv}{\dr}\frac{\drv}{\dr}-\unitvec{e}\right).
\label{eq:squirmer}
\end{equation}
Here, $\drv=\vec{r}-\vec{r}_\text{S}$ is the position vector $\vec{r}$
relative to the squirmer's center $\vec{r}_\text{S}$, $B_1$ and $B_2$
are constants, and $\unitvec{e}$ is the unit orientation vector of the
sphere.

The flow resulting from this boundary condition is governed by the Navier-Stokes equations,
which reduce to the Stokes equations,
\begin{align}
\eta \nabla^2\vec{u}(\vec{r}) &= -\vec{\nabla}p(\vec{r}), \label{eq:stokes-p} \\
\vec{\nabla}\cdot\vec{u}(\vec{r}) &= 0 \label{eq:stokes-m} ,
\end{align}
in the here-applicable limit of a low Reynolds number,
\begin{equation}
\mathrm{Re}=\frac{2\rho v_0 R_\text{S}}{\eta} \ll 1
\label{eq:reynolds}.
\end{equation}
$p$ refers to the pressure, $\rho$ to the fluid density, and $\eta$ to
the fluid viscosity, while $v_0$ is a characteristic flow velocity.
$\vec{\nabla}$, $\vec{\nabla}\cdot$, and $\nabla^2$ are the gradient,
divergence and Laplace operators, respectively.
\Cref{eq:stokes-m,eq:stokes-p} under the condition of
\cref{eq:squirmer} are solved by the flow
field\cite{blake71b,ishikawa06a}
\begin{align}
\vec{u}_\text{S}(\vec{r})=\phantom{+}&
B_1\frac{R_\text{S}^3}{\dr^3}\left(\frac{\unitvec{e}\cdot\drv}{\dr}\frac{\drv}{\dr}-\frac{1}{3}\unitvec{e}\right) \nonumber \\
+& B_2\left(\frac{R_\text{S}^4}{\dr^4}-\frac{R_\text{S}^2}{\dr^2}\right)\left(\frac{3}{2}\left(\frac{\unitvec{e}\cdot\drv}{\dr}\right)^2-\frac{1}{2}\right)\frac{\drv}{\dr} \nonumber \\
+& B_2\frac{R_\text{S}^4}{\dr^4}\frac{\unitvec{e}\cdot\drv}{\dr}\left(\frac{\unitvec{e}\cdot\drv}{\dr}\frac{\drv}{\dr}-\unitvec{e}\right)
\label{eq:squirmerflow}
\end{align}
in the laboratory frame.
This corresponds to the squirmer moving with a velocity\cite{blake71b,ishikawa06a} of
\begin{equation}
\vec{v}_0=\frac{2}{3}B_1\unitvec{e}
\label{eq:squirmer-speed},
\end{equation}
i.e., a velocity that depends only on the
first mode and points in the direction of the squirmer's orientation
vector $\unitvec{e}$.  Micrometer-sized swimmers in water exist in the
low-$\mathrm{Re}$ limit according to \cref{eq:reynolds}, thus $v_0$
only sets the time scale without changing the physical behavior.  This
makes it convenient to scale out $v_0$ and introduce the dipolarity
\begin{equation}
\beta=\frac{B_2}{B_1}
\end{equation}
as the ratio of the magnitudes of the second and first moment.
$\beta$ classifies the shape of the flow field, with the sign
distinguishing three different kinds of swimmers.  A pusher with
$\beta<0$ pushes fluid away from its front and back (with
$\unitvec{e}$ pointing forward) and draws fluid in from its sides.  A
puller with $\beta>0$ pulls fluid toward itself at front and back,
pushing it away from its sides.  At the transition point $\beta=0$
lies the neutral squirmer, which moves fluid from front to back.
Biological examples of these three classes include \emph{Escheria
  coli}\cite{drescher11a}, \emph{Chlamydomonas
  reinhardtii}\cite{drescher10a}, and
\emph{Paramecium}\cite{ishikawa06b}, respectively.

\subsection{Mode Decomposition}
\label{subsec:modes}

We first consider a Stokeslet\cite{stokes51a}, the flow due to a force
monopole $\vec{F}$ applied at $\vec{r}_\text{S}$:
\begin{equation}
	\vec{u}_\text{FM}(\vec{r},\vec{r}_\text{S})=\frac{1}{8\pi\eta}\mathcal{M}(\vec{r},\vec{r}_\text{S}) \vec{F} \label{eq:stokeslet}
\end{equation}
with the Oseen tensor
\begin{equation}
	\mathcal{M}(\vec{r},\vec{r}_\text{S})=\frac{1}{\left|\vec{r}-\vec{r}_\text{S}\right|}\left(
	\mathbb{1}+\frac{(\vec{r}-\vec{r}_\text{S})\otimes(\vec{r}-\vec{r}_\text{S})}{\left|\vec{r}-\vec{r}_\text{S}\right|^2}
	\right)
	.
\end{equation}
We know that the flows of the higher hydrodynamic moments can be
obtained from the Stokeslet \cref{eq:stokeslet} by
differentiation\cite{chwang75a,pozrikidis92a,spagnolie12a}:
\begin{align}
	\vec{u}_\text{FD}(\vec{r},\vec{r}_\text{S})&=\mathcal{D}(\vec{r},\vec{r}_\text{S})\vec{F} \label{eq:forcedipole} \\
	\vec{u}_\text{SD}(\vec{r},\vec{r}_\text{S})&=-\frac{1}{2}\nabla^2_\text{S}\vec{u}_\text{FM}(\vec{r},\vec{r}_\text{S}) \label{eq:sourcedipole} \\
	\vec{u}_\text{SQ}(\vec{r},\vec{r}_\text{S})&=\mathcal{Q}(\vec{r},\vec{r}_\text{S})\vec{F} \label{eq:sourcequadrupole} \\
	\intertext{with}
	\mathcal{D}(\vec{r},\vec{r}_\text{S})&=-\left(\vec{\nabla}_\text{S}\otimes\vec{u}_\text{FM}(\vec{r},\vec{r}_\text{S})\right)^\intercal \\
	\mathcal{Q}(\vec{r},\vec{r}_\text{S})&=\frac{1}{3}\left(\vec{\nabla}_\text{S}\otimes\vec{u}_\text{SD}(\vec{r},\vec{r}_\text{S})\right)^\intercal.
\end{align}
Here, $\otimes$ is the dyadic product and the subscript $_\text{S}$ to differentiation with respect to $\vec{r}_\text{S}$.

\begin{figure}[htb]
\centering
\input{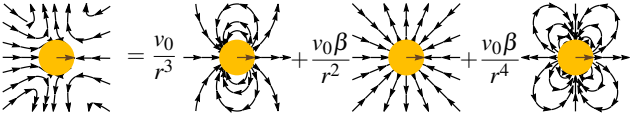}
\caption{
The squirmer's flow field is comprised of a source dipole, a force dipole, and a source quadrupole.
}
\label{fig:moments}
\end{figure}

We can write the squirmer flow in terms of these moments by
identifying them in \cref{eq:squirmerflow}:
\begin{align}
	\vec{u}_\text{S}(\vec{r})
	=\phantom{+}&
	\frac{8\pi\eta}{3}B_1R_\text{S}^3 \vec{u}_\text{SD}(\vec{r},\vec{r}_\text{S})
	+\frac{8\pi\eta}{2}B_2R_\text{S}^2 \vec{u}_\text{FD}(\vec{r},\vec{r}_\text{S}) \nonumber \\
	+&\frac{8\pi\eta}{2}B_2R_\text{S}^4 \vec{u}_\text{SQ}(\vec{r},\vec{r}_\text{S})
	\label{eq:squirmer-decomposition}
\end{align}
with $\vec{F}=\unitvec{e}$.
This decomposition is illustrated by \cref{fig:moments}: a squirmer's flow field is
composed of a source dipole with $r^{-3}$ decay and the prefactor
$B_1$, a force dipole with $r^{-2}$ decay and a prefactor of $B_2$,
and a source quadrupole with $r^{-4}$ decay and the same $B_2$
prefactor.

\subsection{Method of Reflections}
\label{subsec:reflect}

The expressions of \cref{subsec:squirmer} are valid in bulk only.  The
presence of a solid boundary can be incorporated via the method of
reflections\cite{felderhof77b}.  Here, a virtual flow originating from
inside the obstacle is introduced.  Its purpose is to ensure
fulfillment of the no-slip boundary condition on the obstacle's
surface,
\begin{equation}
	\left.\vec{u}(\vec{r})\right|_\text{boundary}= \vec{0}
	\label{eq:noslip}
	.
\end{equation}
We can limit ourselves to obtaining the image of the Stokeslet
$\vec{u}_\text{FM}(\vec{r},\vec{r}_\text{S})$ as \cref{subsec:modes}
permits to express any higher hydrodynamic modes in terms of a Stokeslet.  For a
flat wall, this leads to the \citeauthor{blake71a} tensor, a
superposition of force monopole, force dipole and source
dipole\cite{blake71a,vonhansen11a},
\begin{align}
\mathcal{R}_\text{wall}\vec{u}_\text{FM}(\vec{r},\vec{r}_\text{S})
=&- \vec{u}_\text{FM}(\vec{r}-\vec{r}_*) + \vec{u}_\text{FD}(\vec{r}-\vec{r}_*) \nonumber\\
&- \vec{u}_\text{SD}(\vec{r}-\vec{r}_*)
\label{eq:reflection-wall}.
\end{align}
These seemingly originate from a point $\vec{r}_*$ on the other side
of the wall and at the same distance from the wall as the Stokeslet.
The image Stokeslet for a no-slip sphere
is\cite{oseen27a,spagnolie15a}
\begin{align}
\mathcal{R}\vec{u}_\text{FM}(\vec{r},\vec{r}_\text{S})
&= \frac{1}{8\pi\eta}\mathcal{R}\mathcal{M}(\vec{r},\vec{r}_\text{S})\vec{F}
\label{eq:reflection}
\end{align}
with the reflection operator $\mathcal{R}$ given in
\cref{app:reflection}.  \Cref{eq:reflection-wall} for the reflection
by a flat wall is recovered from this by expanding around $R^{-1}=0$.
The expansion reveals that corrections are to leading order linear in
the inverse radius.

We finally obtain the image squirmer,
\begin{align}
\mathcal{R}
\vec{u}_\text{S}(\vec{r})
	=
	\phantom{+}&\frac{8\pi\eta}{3}B_1R_\text{S}^3 \mathcal{R}\vec{u}_\text{SD}(\vec{r},\vec{r}_\text{S}) \nonumber \\
	+&\frac{8\pi\eta}{2}B_2R_\text{S}^2 \mathcal{R}\vec{u}_\text{FD}(\vec{r},\vec{r}_\text{S}) \nonumber \\
	+&\frac{8\pi\eta}{2}B_2R_\text{S}^4 \mathcal{R}\vec{u}_\text{SQ}(\vec{r},\vec{r}_\text{S}).
	\label{eq:reflected-squirmer-decomposition-1}
\intertext{One can exploit the linearity of the Stokes equation to change the order of operations.
That is, we first apply the known reflection of \cref{eq:reflection} and then perform the differentiation of \cref{eq:forcedipole,eq:sourcedipole,eq:sourcequadrupole}:}
    =
    -&\frac{8\pi\eta}{6}B_1R_\text{S}^3\nabla_\text{S}^2 \left(\mathcal{R}\vec{u}_\text{FM}(\vec{r},\vec{r}_\text{S})\right) \nonumber \\
    -&\frac{8\pi\eta}{2}B_2R_\text{S}^2\left(\vec{\nabla}_\text{S}\otimes\left(\mathcal{R}\vec{u}_\text{FM}(\vec{r},\vec{r}_\text{S})\right)\right)\vec{F} \nonumber \\
    -&\pi\eta B_2R_\text{S}^4\left(\vec{\nabla}_\text{S}\otimes\nabla_\text{S}^2\left(\mathcal{R}\vec{u}_\text{FM}(\vec{r},\vec{r}_\text{S})\right)\right)\vec{F}
    \label{eq:reflected-squirmer-decomposition-2}
    .
\end{align}
The flow field
\begin{equation}
	\vec{u}(\vec{r})=\vec{u}_\text{S}(\vec{r})+\mathcal{R}\vec{u}_\text{S}(\vec{r})	
	\label{eq:fullflow}
\end{equation}
fulfills the Stokes \cref{eq:stokes-m,eq:stokes-p} and the
no-slip boundary condition \cref{eq:noslip} on the obstacle.  It does
not, however, exactly fulfill the slip boundary condition
\cref{eq:squirmer} on the squirmer as this is not possible with a
single reflection.  An infinite series of reflections at the
obstacle's and squirmer's surfaces would be required to respect both
boundary conditions simultaneously.

Other methods that fulfill the condition of \cref{eq:squirmer} by
correctly incorporating near-field hydrodynamic effects are, however,
computationally much more expensive.  The tradeoff of using a
far-field method for a problem that is potentially
near-field-dependent will be justified by \cref{subsec:higherorder},
where we compare some results to ones obtained with a
near-field-capable method.

\subsection{Fax\'en's law}
\label{subsec:faxen}

Now that we have the flow field, we can calculate the squirmer's response to it via Fax\'en's laws\cite{faxen22b,brenner64a,batchelor72b,durlofsky87a,ishikawa06a}.
The first law states that a force-free sphere at position $\vec{r}$ moves with velocity
\begin{equation}
\vec{v}=\left(1+\frac{R_\text{S}^2}{6}\nabla^2\right)\vec{u}(\vec{r}).
\label{eq:faxen-v}
\end{equation}
The second law gives the angular velocity of the sphere as
\begin{equation}
\vec{\omega}=\frac{1}{2}\vec{\nabla}\times\vec{u}(\vec{r}).
\label{eq:faxen-omega}
\end{equation}
\Cref{eq:faxen-v,eq:faxen-omega} can be used to calculate the response
of the squirmer to the reflected flow $\mathcal{R}\vec{u}_\text{S}$.
Its response to $\vec{u}_\text{S}$ cannot be calculated this way as
the flow diverges at $\vec{r}_\text{S}$; however, we already know from
\cref{eq:squirmer-speed} that $\vec{u}_\text{S}$ makes the squirmer
move with velocity $v_0\unitvec{e}$.  While
\cref{eq:faxen-v,eq:faxen-omega} are series expansions, all higher
orders are zero for spheres in Stokes flow\cite{brenner64a}.
Fax\'en's third\cite{batchelor72b} and higher\cite{puljiz19a} laws
are not needed as the squirmer is assumed to be rigid.

\subsection{Numerical Method}
\label{subsec:numerical}

Thus far, we have only given the analytical expressions for the
problem considered.  We now resort to a simple numerical method to
solve the associated equation system, as analytical solutions are not
available.  This requires choosing values for the free parameters,
which are the starting position and orientation of the squirmer and
the squirmer radius $R_\text{S}$ and dipolarity $\beta$.  Due to the
symmetry of the problem, we can restrict ourselves to the $z=0$ plane,
while still considering the full three-dimensional problem.
We can furthermore set $R_\text{S}=1$ without loss of generality as
there are no externally-defined length scales.

The flow field $\vec{u}(\vec{r},t)$ due to the squirmer can be
obtained from \cref{eq:fullflow}.  All derivatives here are carried
out analytically.  The linear velocity $\vec{v}(t)$ and angular
velocity $\vec{\omega}(t)$ of the squirmer is obtained from the flow
via Fax\'en's \cref{eq:faxen-v,eq:faxen-omega}.  The derivatives in
these are carried out numerically via two-sided central finite
differences to avoid further increasing the number of terms in the
expression, which is already approaching the limit of what can be
computed efficiently.  An Euler integrator,
\begin{align}
    \vec{r}(t)&=\vec{r}(t-\tgrid)+\vec{v}(t)\tgrid
    \label{eq:euler},
\end{align}
then updates the position of the squirmer
and the entire process is iterated to obtain the trajectory.
The integrator's time step $\tgrid$ is not relevant as low-Reynolds flow is time-independent;
instead, the squirmer's $v_0$ determines the integrator's step size.
In the calculations below, we adaptively set
\begin{equation}
\tgrid = 0.01\frac{1}{v_0}\max(h, R_\text{S})
.
\end{equation}
This allows for fast integration of the trajectory far away from the
obstacle, where the squirmer moves in a (nearly) straight line, and a
high resolution when the gap between the squirmer and the obstacle is
small and the hydrodynamic interactions are strong.

We include a hard-core repulsive potential that prevents the overlap
of squirmer and obstacle by ensuring that $h\geq r_\text{cut}$.  This
modifies the integrator of \cref{eq:euler} to become
\begin{align}
\vec{r}(t)&=
\max\left(r_\text{cut}+R+R_\text{S},
r^\prime(t)
\right)
\unitvec{r}^\prime(t)
\intertext{with}
\vec{r}^\prime(t)&=\vec{r}(t-\tgrid)+\vec{v}(t)\tgrid
.
\end{align}

\subsection{Stability Analysis using Lubrication Theory}
\label{subsec:lubrication}

The flow in a small gap between two objects is the regime of
lubrication theory\cite{goldman67a,cichocki98a,ishikawa06a}.  It
assumes that the flow is dominated by the interaction between those
points where the surfaces are closest.  For a squirmer near a flat
wall, \citet{lintuvuori16a} give
\begin{align}
\frac{\mathrm{d}\varphi}{\mathrm{d}t}=
\vec{\omega} &=\frac{3v_0}{2R_\text{S}}\cos{\varphi}\left(1-\beta\sin\varphi\right) \unitvec{e}_\varphi+\mathcal{O}\left(1/\log \frac{h}{R_\text{S}}\right) \\
\vec{v}&=0+\mathcal{O}\left(1/\log \frac{h}{R_\text{S}}\right)
\end{align}
where $\unitvec{e}_\varphi$ is the angular unit vector in our angle convention.
This means that lateral translation vanishes but rotation remains possible.

Solving the above equation for the stationary state
$\omega = 0$ yields one stable solution,
$\varphi=-90^\circ$ at $\beta<-1$.  This corresponds to the hovering
state of a pusher.  However, the interplay between the lubrication
regime and far-field reorientation can lead to orbiting states of both
the puller and pusher\cite{lintuvuori16a}.

Note that lubrication theory is an extreme limit where far-field
hydrodynamics become irrelevant and only one term at infinitely close
separation remains of the near-field flow.  The typical gap sizes $h$
we will find in the next section fall in between those where far-field
hydrodynamics is applicable $(h\gtrsim R_\text{S}$) and those where
lubrication theory is valid $(\log ( h /R_\text{S}) \ll -1$).  This
necessitates verification of such predictions using methods that deal
with the intermediate regime, the near field.

\subsection{Resolving the Near Field using Lattice Boltzmann}
\label{subsec:LB}

Neither the far-field calculations of
\cref{subsec:reflect,subsec:faxen,subsec:numerical} nor the
lubrication considerations of \cref{subsec:lubrication} are able to
accurately capture the intermediate near-field regime.  The squirmer
enters this regime when it comes close to the obstacle, and we
therefore resort to the LB method\cite{mcnamara88a,higuera89a} to test
our far-field predictions.  LB is a Navier-Stokes solver that
excels at coupled fluid-particle simulations and
flows in complex geometries\cite{ladd94a}.
Space and time are discretized on a grid of spacing $\Delta x$ and
$\Delta t$, respectively, and the Boltzmann transport equation is
solved using a two-time relaxation scheme. This reproduces solutions to
the Navier-Stokes equations on sufficiently large length and time
scales, which we resolve.  Our specific implementation using the
waLBerla framework\cite{godenschwager13a,rettinger17a} and its
application to the present problem are described in
Ref.~\citenum{kuron19a}.  We refer the interested reader to
Ref.~\citenum{krueger17a} for a complete overview of the general LB
method.

We should point out the following concerning our LB calculations here:
The minimal gap between squirmer and obstacle that LB can accurately
resolve is limited to around $\Delta x$, due to our LB's lack of
lubrication corrections\cite{durlofsky87a,nguyen02b,ishikawa08b}.
Note that such corrections exist for driven spheres and some other shapes,
but a specific implementation for a squirmer has not yet been
formulated, due to the complexity of the boundary problem.  We
therefore impose short-ranged Weeks-Chandler-Andersen-type
potential\cite{weeks71a} between the obstacle and squirmer, which is
given by
\begin{equation}
    U(\vec{r}_\text{S}-\vec{r}) = 
        4 \epsilon \left[\left(\frac{\sigma}{\left|\vec{r}_\text{S} - \vec{r}\right|}\right)^{12} - \left(\frac{\sigma}{\left|\vec{r}_\text{S} - \vec{r}\right|}\right)^{6} \right] + \epsilon \text{ ,}
\end{equation}
for $0 < |\vec{r}_\text{S}| < 2^{1/6}\sigma$ and set to 0 for
$ r > 2^{1/6}\sigma$.  This potential ensures that the squirmer and
obstacle remain separated by at least one LB cell.  We do not resort
to hard-core repulsions, as a discontinuous potential leads to issues
with the underlying algorithm for the positional update of our
squirmer in waLBerla\cite{godenschwager13a,rettinger17a}.

A resolution of eight lattice cells per squirmer radius
($R_\text{S}=8\Delta x$) was used throughout, which allows for
accurate capturing of flow fields down to gaps of
$h\gtrsim R_\text{S}/8$, also see Ref.~\citenum{kuron19a}.  The
calculations in this paper further employ a periodic calculation
domain of size $L\times L\times H$ with $L= \max(5.5R, 160 \Delta x)$
and $H= \max(2.6R, 80 \Delta x)$.  The viscosity is set to
$\eta= 0.8\rho\Delta x^2/\Delta t$.  The squirmer is initially located
at $(\max(0.3R, 9 \Delta x),y_0,L/2)^\intercal$, while the obstacle is at
$(L/2,L/2,H/2)^\intercal$.  In all LB calculations,
$\epsilon=1\rho\Delta x^5/\Delta t^2$ and $\sigma=1.34\Delta x$ are
used.

\section{Characterization}
\label{sec:characterization}

We start our analysis of the behavior of a squirmer near a spherical
obstacle using the far-field hydrodynamic theory of
\cref{subsec:reflect,subsec:faxen,subsec:numerical}.  To determine the
different behaviors, we vary the three free parameters in the model:
the relative obstacle size $R/R_\text{S}$, the squirmer dipolarity
$\beta$, and the initial off-axis position $y_0$ (or equivalently, the
initial incidence angle $\varphi_0$).  We pick
$R/R_\text{S}\in\left\{1,2,5,10,20,50,100,200,500\right\}$ and
$y_0\in\left[0,10R\right]$ spaced roughly exponentially,
$\beta\in\left[-30,30\right]$ spaced linearly, and
$r_\text{cut}/R_\text{S}\in\left\{0,0.01,0.1,0.2\right\}$.

\begin{figure}[!tb]
\centering
\input{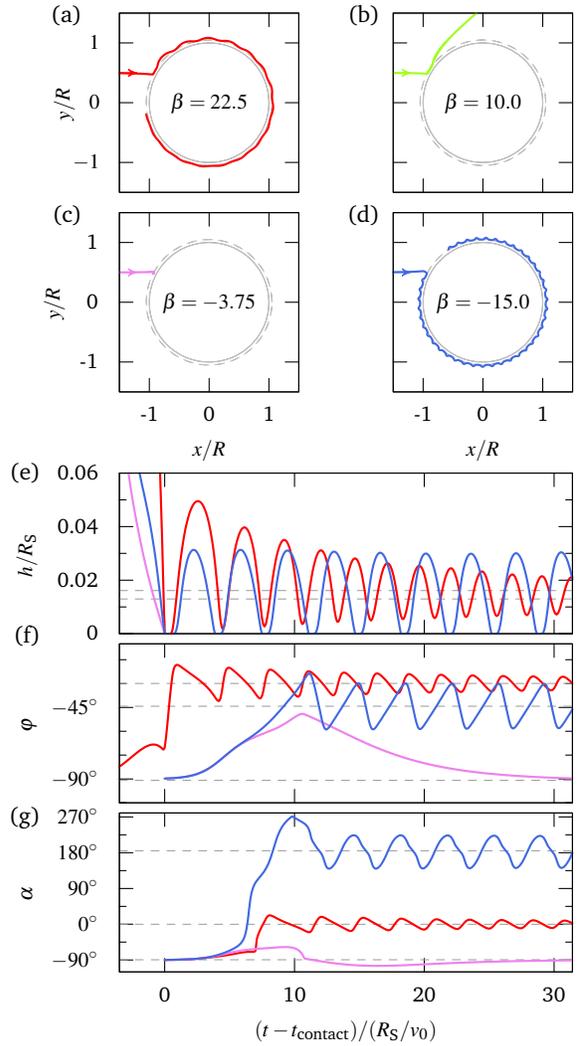} 
\caption{ (a-d) Trajectories for different $\beta$, leading to a
  forward orbit (red, $\beta=22.5$), scattering (green, $\beta=10$), a
  hovering state (violet, $\beta=-3.75$) and a backward orbit (blue,
  $\beta=-15$).  The solid gray circle indicates the position of the
  obstacle, while the dashed gray circle indicates the closest
  possible approach.  (e) The surface-to-surface gap size
  $h/R_\text{S}$ between squirmer and obstacle over time.  The dashed
  gray lines indicate the long-time mean gap size.  (f) The angle
  $\varphi$ between the obstacle's surface and the squirmer's
  orientation vector over time.  (g) The angle $\alpha$ between the
  obstacle's surface and the squirmer's direction of motion over time.
  The dashed gray lines indicate the long-time mean angle.  All
  calculations used $R=20R_\text{S}$, $y_0=0.5R$, $r_\text{cut}=0$.
  Time $t_\text{contact}$ is the time at which the squirmer first made
  contact ($h=r_\text{cut}$) with the obstacle.  }
\label{fig:traj}
\end{figure}

\subsection{Trajectories}
\label{subsec:trajectories}

Inspecting the resulting trajectories reveals four general classes of behavior,
examples of which are shown in \cref{fig:traj}a-d.
(a) corresponds to a forward orbit;
(b) is a scattering trajectory;
in (c) the squirmer hovers above obstacle's surface;
(d) is a backward orbit.
\Cref{fig:traj}e shows the size of the gap between the squirmer and the obstacle over time.
Here, one can see that a decaying oscillation is modulated onto the forward orbit,
while the oscillation of the backward orbit is quite stable.
These oscilations are reminiscent of those observed in Refs.~\citenum{ishimoto13a,lintuvuori16a}.
\Cref{fig:traj}fg shows the orientation of the squirmer's orientation vector $\unitvec{e}$ and velocity vector $\vec{v}$, respectively. 

\begin{figure}[tb]
\centering
\begin{tikzpicture}
\usetikzlibrary{calc}
\usetikzlibrary{patterns}

\tikzset{vector/.style={->,line width=1pt}}
\tikzset{object/.style={line width=1pt}}
\tikzset{obstacle/.style={pattern=north west lines}}

\pgfmathsetmacro\Rs{0.5}
\pgfmathsetmacro\gap{3*\Rs}
\pgfmathsetmacro\L{3*\Rs+1.5*\gap}
\pgfmathsetmacro\yoff{2.3*\Rs}

\input{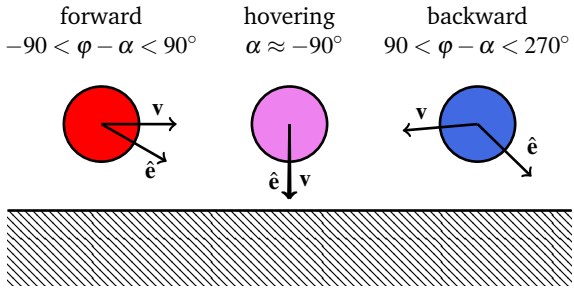}

\draw[object] (-\L,0) -- (\L,0);
\draw[obstacle,draw=none] (-\L,0) -- (\L,0) -- (\L,-1) -- (-\L,-1) -- (-\L,0);

\begin{scope}[shift={(-2*\Rs-\gap,\yoff)}]
	\pgfmathsetmacro\eang{\eangf}
	\pgfmathsetmacro\vang{\vangf}
	\def\piclabel{forward \\ $-90<\varphi-\alpha<90^\circ$}
	
	\definecolor{gpred}{rgb}{1,0,0}
	\draw[object,fill=gpred] (0,0) circle(\Rs) node[above=\Rs*1.5cm,align=center] {\piclabel};
	
	\draw [vector] (0,0) -- node[below,pos=0.75] {$\unitvec{e}$} ++(\eang:2*\Rs);
	
	\draw [vector] (0,0) -- node[above,pos=0.75] {$\vec{v}$} ++(\vang:2*\Rs);
\end{scope}

\begin{scope}[shift={(2*\Rs+\gap,\yoff)}]
	\pgfmathsetmacro\eang{\eangb}
	\pgfmathsetmacro\vang{\vangb}
	\def\piclabel{backward \\ $90<\varphi-\alpha<270^\circ$}
	
	\definecolor{gpblue}{rgb}{0.255,0.412,0.882}
	\draw[object,fill=gpblue] (0,0) circle(\Rs) node[above=\Rs*1.5cm,align=center] {\piclabel};
	
	\draw [vector] (0,0) -- node[above right,pos=0.75] {$\unitvec{e}$} ++(\eang:2*\Rs);
	
	\draw [vector] (0,0) -- node[above,pos=0.75] {$\vec{v}$} ++(\vang:2*\Rs);
\end{scope}

\begin{scope}[shift={(0,\yoff)}]
	\pgfmathsetmacro\eang{\eangs}
	\pgfmathsetmacro\vang{\vangs}
	\def\piclabel{hovering \\ $\phantom{\left|\right.}\alpha\approx-90^\circ$}
	
	\definecolor{gpviolet}{rgb}{0.933,0.510,0.933}
	\draw[object,fill=gpviolet] (0,0) circle(\Rs) node[above=\Rs*1.5cm,align=center] {\piclabel};
	
	\draw [vector] (0,0) -- node[left,pos=0.75] {$\unitvec{e}$} ++(\eang:2*\Rs);
	
	\draw [vector] (0,0) -- node[right,pos=0.75] {$\vec{v}$} ++(\vang:2*\Rs);
\end{scope}

\end{tikzpicture}
\caption{ The bound states correspond to characteristic values of the
  angles introduces in \cref{fig:angles}.  Forward orbiting/sliding
  corresponds to $\vec{v}$ and $\unitvec{e}$ approximately parallel,
  backward orbiting/sliding to $\vec{v}$ and $\unitvec{e}$
  approximately antiparallel, and hovering is $\vec{v}$ pointing
  approximately perpendicularly into the obstacle.  The angles
  illustrated here correspond to the dashed lines in
  \cref{fig:traj}fg.  }
\label{fig:angle-cases}
\end{figure}

To identify the origin of the three different kinds of bound
trajectories more clearly, these angles are illustrated in
\cref{fig:angle-cases}: In a forward orbit, $\unitvec{e}$ and
$\vec{v}$ are roughly parallel ($\unitvec{e}\cdot\vec{v}>0$), while in
a backward orbit they are antiparallel ($\unitvec{e}\cdot\vec{v}<0$).
Hovering is a case where $\vec{v}$ points almost straight into the
obstacle, so the squirmer is stuck in place.  A squirmer is considered
hovering when it moves at a speed of $v<v_0/100$ or has
$\left|\alpha-\varphi\right|<3^\circ$.  The precise choice of these
limits may appear arbitrary, but we found that most orbiting
trajectories exhibit angles that either much larger or much smaller.
If the trajectory is oscillatory, we average the angle over at least one orbit.

Lastly, we should note that all of the above classes of trajectories
are also obtained in our LB calculations of \cref{subsec:LB}.
However, the specific parameters for which these behaviors are
observed are different.  We will come back to this in
\cref{subsec:higherorder}.

\subsection{Interpreting the Bound States}
\label{subsec:interpretation}

\begin{figure}[!tb]
\centering
\input{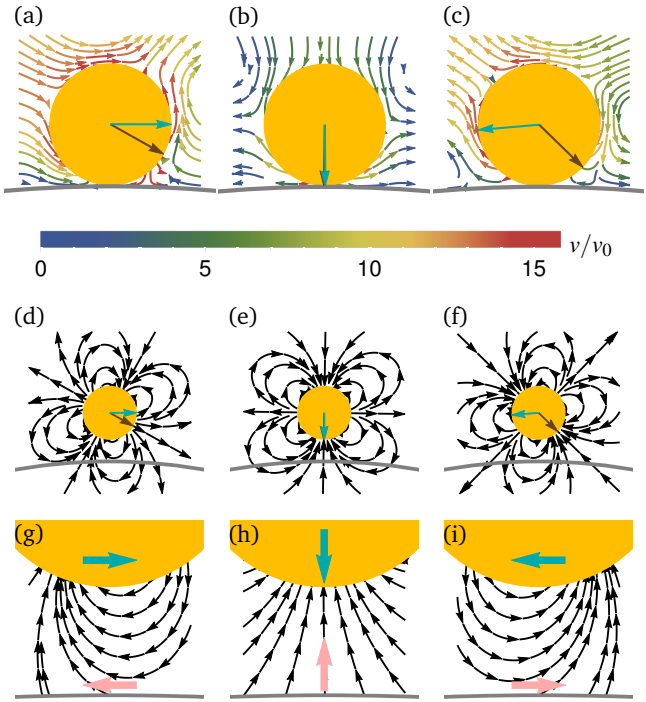}
\caption{ The flow around the squirmer in the (a) forward (puller),
  (b) hovering (pusher), and (c) backward (pusher) configurations of
  \cref{fig:angle-cases}.  The cyan arrow indicates $\vec{v}$, while
  the brown one marks $\unitvec{e}$.  The gray line indicates the
  position of the no-slip wall.  (d-f) The same configurations, but
  showing the bulk flow due to the source quadrupole moment only, where we have not accounted for the boundary condition. (g-i) Zoom-ins on the gap in (d-f), showing
  that the source quadrupole leads to a flow at the surface (closest point)
  corresponding to the pink arrow. 
  The closest point dominates the dynamics and to satisfy the boundary condition, the swimmer effectively moves with $\vec{v}$ as indicated by the cyan arrow.}
\label{fig:flow}
\end{figure}

\begin{figure}[htb]
\centering
\begin{tikzpicture}[gnuplot]
\tikzset{every node/.append style={scale=0.88}}
\path (0.000,0.000) rectangle (8.000,4.000);
\gpcolor{color=gp lt color border}
\gpsetlinetype{gp lt border}
\gpsetdashtype{gp dt solid}
\gpsetlinewidth{1.00}
\draw[gp path] (1.155,0.540)--(1.335,0.540);
\draw[gp path] (3.516,0.540)--(3.336,0.540);
\node[gp node right] at (0.994,0.540) {$-15$};
\draw[gp path] (1.155,1.000)--(1.335,1.000);
\draw[gp path] (3.516,1.000)--(3.336,1.000);
\node[gp node right] at (0.994,1.000) {$-10$};
\draw[gp path] (1.155,1.460)--(1.335,1.460);
\draw[gp path] (3.516,1.460)--(3.336,1.460);
\node[gp node right] at (0.994,1.460) {$-5$};
\draw[gp path] (1.155,1.920)--(1.335,1.920);
\draw[gp path] (3.516,1.920)--(3.336,1.920);
\node[gp node right] at (0.994,1.920) {$0$};
\draw[gp path] (1.155,2.379)--(1.335,2.379);
\draw[gp path] (3.516,2.379)--(3.336,2.379);
\node[gp node right] at (0.994,2.379) {$5$};
\draw[gp path] (1.155,2.839)--(1.335,2.839);
\draw[gp path] (3.516,2.839)--(3.336,2.839);
\node[gp node right] at (0.994,2.839) {$10$};
\draw[gp path] (1.155,3.299)--(1.335,3.299);
\draw[gp path] (3.516,3.299)--(3.336,3.299);
\node[gp node right] at (0.994,3.299) {$15$};
\node[gp node center] at (1.598,0.270) {forw.};
\node[gp node center] at (2.336,0.270) {hov.};
\node[gp node center] at (3.073,0.270) {backw.};
\draw[gp path] (1.155,3.299)--(1.155,0.540)--(3.516,0.540)--(3.516,3.299)--cycle;
\node[gp node center,rotate=-270] at (0.241,1.919) {$u_\parallel/v_0$, forward/backward};
\node[gp node right] at (2.854,3.864) {total};
\def\gpfillpath{(3.015,3.797)--(3.839,3.797)--(3.839,3.932)--(3.015,3.932)--cycle}
\gpfill{color=gpbgfillcolor} \gpfillpath;
\gpfill{rgb color={0.753,0.753,0.753},gp pattern 1,pattern color=.} \gpfillpath;
\def\gpfillpath{(1.266,0.608)--(1.931,0.608)--(1.931,1.921)--(1.266,1.921)--cycle}
\gpfill{color=gpbgfillcolor} \gpfillpath;
\gpfill{rgb color={0.753,0.753,0.753},gp pattern 1,pattern color=.} \gpfillpath;
\def\gpfillpath{(2.003,1.912)--(2.669,1.912)--(2.669,1.921)--(2.003,1.921)--cycle}
\gpfill{color=gpbgfillcolor} \gpfillpath;
\gpfill{rgb color={0.753,0.753,0.753},gp pattern 1,pattern color=.} \gpfillpath;
\def\gpfillpath{(2.741,1.920)--(3.406,1.920)--(3.406,2.860)--(2.741,2.860)--cycle}
\gpfill{color=gpbgfillcolor} \gpfillpath;
\gpfill{rgb color={0.753,0.753,0.753},gp pattern 1,pattern color=.} \gpfillpath;
\node[gp node right] at (2.854,3.594) {force dipole};
\gpfill{rgb color={0.580,0.000,0.827}} (3.015,3.527)--(3.839,3.527)--(3.839,3.662)--(3.015,3.662)--cycle;
\gpfill{rgb color={0.580,0.000,0.827}} (1.330,1.920)--(1.497,1.920)--(1.497,1.921)--(1.330,1.921)--cycle;
\gpfill{rgb color={0.580,0.000,0.827}} (2.068,1.920)--(2.235,1.920)--(2.235,1.921)--(2.068,1.921)--cycle;
\gpfill{rgb color={0.580,0.000,0.827}} (2.806,1.920)--(2.973,1.920)--(2.973,1.921)--(2.806,1.921)--cycle;
\node[gp node right] at (6.737,3.864) {source dipole};
\gpfill{rgb color={0.000,0.620,0.451}} (6.898,3.797)--(7.722,3.797)--(7.722,3.932)--(6.898,3.932)--cycle;
\gpfill{rgb color={0.000,0.620,0.451}} (1.515,1.881)--(1.682,1.881)--(1.682,1.921)--(1.515,1.921)--cycle;
\gpfill{rgb color={0.000,0.620,0.451}} (2.252,1.920)--(2.420,1.920)--(2.420,1.921)--(2.252,1.921)--cycle;
\gpfill{rgb color={0.000,0.620,0.451}} (2.990,1.888)--(3.157,1.888)--(3.157,1.921)--(2.990,1.921)--cycle;
\node[gp node right] at (6.737,3.594) {source quadrupole};
\gpfill{rgb color={0.337,0.706,0.914}} (6.898,3.527)--(7.722,3.527)--(7.722,3.662)--(6.898,3.662)--cycle;
\gpfill{rgb color={0.337,0.706,0.914}} (1.699,0.646)--(1.866,0.646)--(1.866,1.921)--(1.699,1.921)--cycle;
\gpfill{rgb color={0.337,0.706,0.914}} (2.437,1.912)--(2.604,1.912)--(2.604,1.921)--(2.437,1.921)--cycle;
\gpfill{rgb color={0.337,0.706,0.914}} (3.175,1.920)--(3.342,1.920)--(3.342,2.891)--(3.175,2.891)--cycle;
\draw[gp path] (1.155,0.540)--(1.335,0.540);
\draw[gp path] (3.516,0.540)--(3.336,0.540);
\node[gp node right] at (0.994,0.540) {$-15$};
\draw[gp path] (1.155,1.000)--(1.335,1.000);
\draw[gp path] (3.516,1.000)--(3.336,1.000);
\node[gp node right] at (0.994,1.000) {$-10$};
\draw[gp path] (1.155,1.460)--(1.335,1.460);
\draw[gp path] (3.516,1.460)--(3.336,1.460);
\node[gp node right] at (0.994,1.460) {$-5$};
\draw[gp path] (1.155,1.920)--(1.335,1.920);
\draw[gp path] (3.516,1.920)--(3.336,1.920);
\node[gp node right] at (0.994,1.920) {$0$};
\draw[gp path] (1.155,2.379)--(1.335,2.379);
\draw[gp path] (3.516,2.379)--(3.336,2.379);
\node[gp node right] at (0.994,2.379) {$5$};
\draw[gp path] (1.155,2.839)--(1.335,2.839);
\draw[gp path] (3.516,2.839)--(3.336,2.839);
\node[gp node right] at (0.994,2.839) {$10$};
\draw[gp path] (1.155,3.299)--(1.335,3.299);
\draw[gp path] (3.516,3.299)--(3.336,3.299);
\node[gp node right] at (0.994,3.299) {$15$};
\node[gp node center] at (1.598,0.270) {forw.};
\node[gp node center] at (2.336,0.270) {hov.};
\node[gp node center] at (3.073,0.270) {backw.};
\draw[gp path] (1.155,3.299)--(1.155,0.540)--(3.516,0.540)--(3.516,3.299)--cycle;
\gpdefrectangularnode{gp plot 1}{\pgfpoint{1.155cm}{0.540cm}}{\pgfpoint{3.516cm}{3.299cm}}
\draw[gp path] (4.994,0.540)--(5.174,0.540);
\draw[gp path] (7.516,0.540)--(7.336,0.540);
\node[gp node right] at (4.833,0.540) {$-6$};
\draw[gp path] (4.994,1.000)--(5.174,1.000);
\draw[gp path] (7.516,1.000)--(7.336,1.000);
\node[gp node right] at (4.833,1.000) {$-4$};
\draw[gp path] (4.994,1.460)--(5.174,1.460);
\draw[gp path] (7.516,1.460)--(7.336,1.460);
\node[gp node right] at (4.833,1.460) {$-2$};
\draw[gp path] (4.994,1.920)--(5.174,1.920);
\draw[gp path] (7.516,1.920)--(7.336,1.920);
\node[gp node right] at (4.833,1.920) {$0$};
\draw[gp path] (4.994,2.379)--(5.174,2.379);
\draw[gp path] (7.516,2.379)--(7.336,2.379);
\node[gp node right] at (4.833,2.379) {$2$};
\draw[gp path] (4.994,2.839)--(5.174,2.839);
\draw[gp path] (7.516,2.839)--(7.336,2.839);
\node[gp node right] at (4.833,2.839) {$4$};
\draw[gp path] (4.994,3.299)--(5.174,3.299);
\draw[gp path] (7.516,3.299)--(7.336,3.299);
\node[gp node right] at (4.833,3.299) {$6$};
\node[gp node center] at (5.467,0.270) {forw.};
\node[gp node center] at (6.255,0.270) {hov.};
\node[gp node center] at (7.043,0.270) {backw.};
\draw[gp path] (4.994,3.299)--(4.994,0.540)--(7.516,0.540)--(7.516,3.299)--cycle;
\node[gp node center,rotate=-270] at (4.241,1.919) {$u_\perp/v_0$, attraction/repulsion};
\def\gpfillpath{(5.112,1.785)--(5.823,1.785)--(5.823,1.921)--(5.112,1.921)--cycle}
\gpfill{color=gpbgfillcolor} \gpfillpath;
\gpfill{rgb color={0.753,0.753,0.753},gp pattern 1,pattern color=.} \gpfillpath;
\def\gpfillpath{(5.900,1.690)--(6.611,1.690)--(6.611,1.921)--(5.900,1.921)--cycle}
\gpfill{color=gpbgfillcolor} \gpfillpath;
\gpfill{rgb color={0.753,0.753,0.753},gp pattern 1,pattern color=.} \gpfillpath;
\def\gpfillpath{(6.688,1.730)--(7.399,1.730)--(7.399,1.921)--(6.688,1.921)--cycle}
\gpfill{color=gpbgfillcolor} \gpfillpath;
\gpfill{rgb color={0.753,0.753,0.753},gp pattern 1,pattern color=.} \gpfillpath;
\gpfill{rgb color={0.580,0.000,0.827}} (5.181,0.956)--(5.360,0.956)--(5.360,1.921)--(5.181,1.921)--cycle;
\gpfill{rgb color={0.580,0.000,0.827}} (5.969,0.627)--(6.148,0.627)--(6.148,1.921)--(5.969,1.921)--cycle;
\gpfill{rgb color={0.580,0.000,0.827}} (6.757,0.765)--(6.936,0.765)--(6.936,1.921)--(6.757,1.921)--cycle;
\gpfill{rgb color={0.000,0.620,0.451}} (5.378,1.809)--(5.557,1.809)--(5.557,1.921)--(5.378,1.921)--cycle;
\gpfill{rgb color={0.000,0.620,0.451}} (6.166,1.690)--(6.345,1.690)--(6.345,1.921)--(6.166,1.921)--cycle;
\gpfill{rgb color={0.000,0.620,0.451}} (6.954,1.767)--(7.133,1.767)--(7.133,1.921)--(6.954,1.921)--cycle;
\gpfill{rgb color={0.337,0.706,0.914}} (5.575,1.920)--(5.754,1.920)--(5.754,2.859)--(5.575,2.859)--cycle;
\gpfill{rgb color={0.337,0.706,0.914}} (6.363,1.920)--(6.542,1.920)--(6.542,3.213)--(6.363,3.213)--cycle;
\gpfill{rgb color={0.337,0.706,0.914}} (7.151,1.920)--(7.330,1.920)--(7.330,3.039)--(7.151,3.039)--cycle;
\draw[gp path] (4.994,0.540)--(5.174,0.540);
\draw[gp path] (7.516,0.540)--(7.336,0.540);
\node[gp node right] at (4.833,0.540) {$-6$};
\draw[gp path] (4.994,1.000)--(5.174,1.000);
\draw[gp path] (7.516,1.000)--(7.336,1.000);
\node[gp node right] at (4.833,1.000) {$-4$};
\draw[gp path] (4.994,1.460)--(5.174,1.460);
\draw[gp path] (7.516,1.460)--(7.336,1.460);
\node[gp node right] at (4.833,1.460) {$-2$};
\draw[gp path] (4.994,1.920)--(5.174,1.920);
\draw[gp path] (7.516,1.920)--(7.336,1.920);
\node[gp node right] at (4.833,1.920) {$0$};
\draw[gp path] (4.994,2.379)--(5.174,2.379);
\draw[gp path] (7.516,2.379)--(7.336,2.379);
\node[gp node right] at (4.833,2.379) {$2$};
\draw[gp path] (4.994,2.839)--(5.174,2.839);
\draw[gp path] (7.516,2.839)--(7.336,2.839);
\node[gp node right] at (4.833,2.839) {$4$};
\draw[gp path] (4.994,3.299)--(5.174,3.299);
\draw[gp path] (7.516,3.299)--(7.336,3.299);
\node[gp node right] at (4.833,3.299) {$6$};
\node[gp node center] at (5.467,0.270) {forw.};
\node[gp node center] at (6.255,0.270) {hov.};
\node[gp node center] at (7.043,0.270) {backw.};
\draw[gp path] (4.994,3.299)--(4.994,0.540)--(7.516,0.540)--(7.516,3.299)--cycle;
\gpdefrectangularnode{gp plot 2}{\pgfpoint{4.994cm}{0.540cm}}{\pgfpoint{7.516cm}{3.299cm}}
\end{tikzpicture}
\caption{Histogram showing the relative contributions of the hydrodynamic modes to the motion of the squirmer, as measured by the bulk flow generated at the point closest to the obstacle (not accounting for this boundary, see text).
  The parallel $\vec{u}_{\parallel}$ and perpendicular $\vec{u}_{\perp}$ components of this flow velocity---corresponding to the configurations given in~\cref{fig:traj}---are provided in the left- and right-hand panels, respectively. The colored bars indicate the contributions of the three hydrodynamic modes, while the net effect is provided by the shaded area.}
\label{fig:strength}
\end{figure}
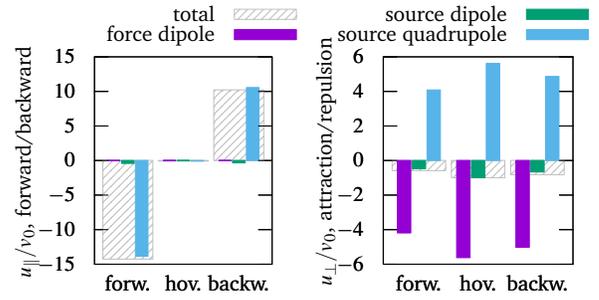

We examine the flow fields in \cref{fig:flow}a-c to identify the hydrodynamic reason behind the observed bound states.
Our far-field analysis relies on Faxen's laws, \cref{eq:faxen-v,eq:faxen-omega}, to implicitly carry out the surface integrals that specify the forces and torques that move and reorient the squirmer in proximity to the boundary.
However, we will argue here that applying the intuitive arguments encountered in lubrication theory, see \cref{subsec:lubrication}, gives insight into the origin of the various bound states.
Lubrication theory states that the dynamics of the squirmer is governed by the viscous dissipation taking place at the point of closest approach\cite{lintuvuori16a}.
In the far field, one can expect a dominant contribution to the surface integral to also come from this point, especially for small separation.

Let us now consider the flow generated by the squirmer at the location of the boundary, without accounting for the boundary's presence.
Thus, we are considering the \textit{unmodified} (bulk) flow field of the squirmer, evaluated at the  point of closest approach.
The \textit{unmodified} fluid velocity at this point is provided in \cref{fig:strength} which decomposes it into components parallel and perpendicular to the boundary.
This figure also shows the separate contributions of the various hydrodynamic modes to the \textit{unmodified} flow field around the squirmer.
Clearly, the source quadrupole moment gives rise to the strongest parallel flow in this scenario.
The perpendicular components of the force dipole and source quadrupole essentially balance, such that \textit{unmodified} flow `into' the wall is dominated by the source dipole. 

The perpendicular component is associated with motion toward/away from the boundary and is thus not of interest here.
Focusing on the boundary-parallel contribution and dominant flow of the source quadrupole, we obtain \textit{unmodified} flow fields due to this term as depicted in \cref{fig:flow}d-f.
Zoom-ins on the region of smallest separation are provided in \cref{fig:flow}g-i, where we should again stress that we only indicate the position of the boundary, but do not account for it in drawing the flow lines.
Clearly, the zero-velocity boundary condition is not satisfied.
To achieve this condition at the point of closest approach, we can assign a velocity to the swimmer that is equal in magnitude, but oppositely directed.
Hence, a forward-moving (\cref{fig:flow}adg), hovering (\cref{fig:flow}beh), and backward-moving (\cref{fig:flow}cfi) state are expected.

The above argument relies on the strong reduction that the dynamics of the squirmer is sufficiently dominated by the point of closest approach, which is only true in the lubrication limit.
In addition, we estimate the contribution there through the flow field around a squirmer in bulk fluid.
The above explanation should thus be seen as a means to develop some intuition for the behavior of the squirmer, but not as a full proof.
\Cref{subsec:higherorder} will show that this intuition is, however, accurate, as removal of the source quadrupole term strongly alters the dynamics.
Similarly, increasing the separation between the squirmer and boundary sufficiently for longer-ranged hydrodynamic modes to dominate also eliminates the backward-orbiting state.
This will be done in \cref{subsec:shortrange} by varying the short-range interaction potential.

\section{Results}
\label{sec:results}

\begin{figure*}[tb]
\centering
\input{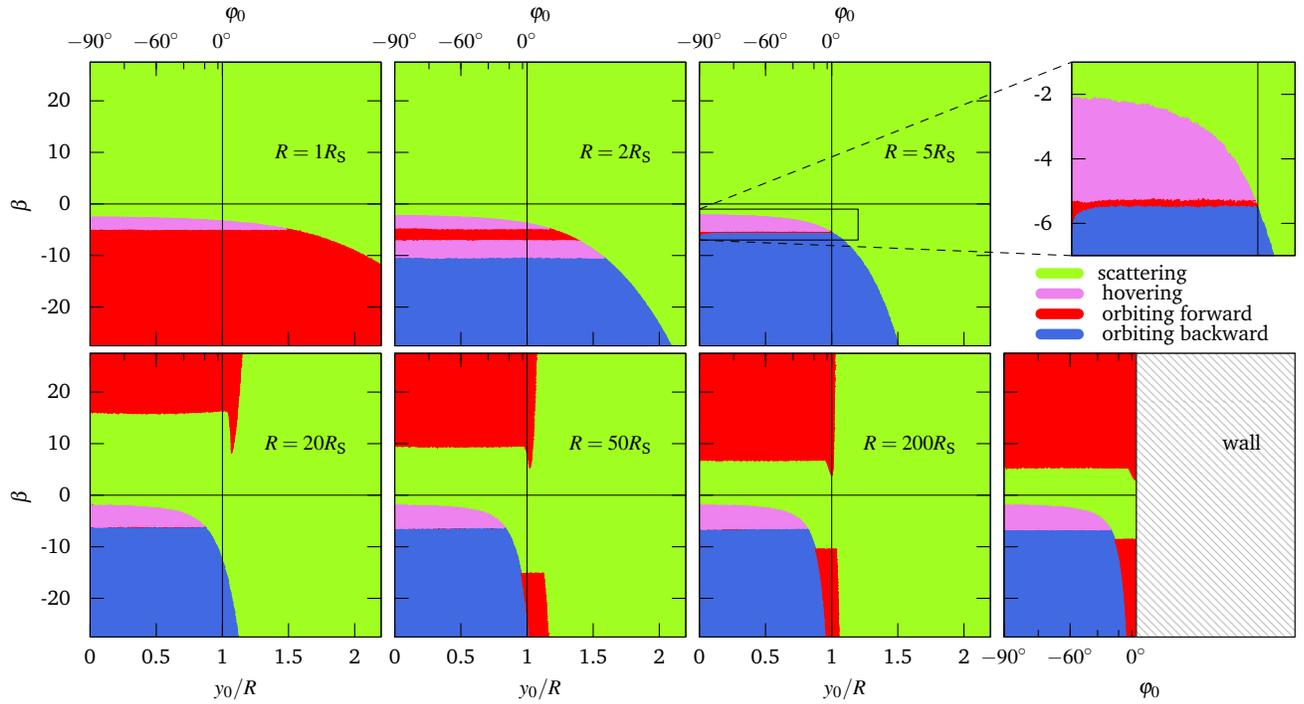}
\caption{ The influence of obstacle-squirmer size ratio: State
  diagrams indicating the way the squirmer behavior depends on
  dipolarity $\beta$ and initial position $y_0$ (or equivalently,
  initial orientation $\varphi_0$) with different $R$ and constant
  $r_\text{cut}=0$.  Forward orbits are red, backward orbits blue,
  hovering is violet and scattering is green, as indicated in the
  legend and used throughout.  The former three (bound) states
  correspond to specific relative angles of the squirmer's orientation
  and motion, which are illustrated in \cref{fig:angle-cases}.  }
\label{fig:vor-R}
\end{figure*}

We are now able to use the observations of \cref{sec:characterization}
to classify trajectories into distinct categories corresponding to the four archetypal trajectories of \cref{fig:traj}.
This allows us to construct state diagrams in parameter space using
the efficient far-field method of
\cref{subsec:reflect,subsec:faxen,subsec:numerical}.  To produce
two-dimensional $\beta$-$y_0$ diagrams, $r_\text{cut}$ and
$R/R_\text{S}$ are chosen to be constant values.  From the data set
discussed in the previous section, one can already obtain coarse state
diagrams by performing a Voronoi construction\cite{aurenhammer91a} to
identify polygonal regions in the two-dimensional parameter space.
Each of these regions is associated with a $(y_0,\beta)$ data point
and contains all points that are closer to this data point than to any
other data point.  The entire region is then filled with the color
assigned to the behavior observed for the respective data point.  We now 
refine the coarse diagram: One can identify polygon
vertices that that connect polygons of different color (i.e., that lie
on the edge of a state's region).  A new calculation is then started
at each of the identified vertices.  This procedure is repeated until
a sufficiently smooth diagram is obtained.

The topology of these state diagrams is roughly as follows.  At
$y_0=0$, only hovering states can be found because the symmetry
remains unbroken.  For nonzero $y_0$, the strongest pushers follow
oscillating backward orbits.  Decreasing the squirmer strength
successively leads to hovering, a forward orbit with decaying
oscillation, and another hovering state, before transitioning to
scattering near $\beta=0$.  Pullers of sufficient strength again enter
into forward orbits with decaying oscillation.  Another region of
forward orbits is found for strong pushers near $y_0=R_\text{S}$, but
with a persistent oscillation.
In the rest of this section, we will discuss various influences on the state diagram:
\hyperref[subsec:obstaclesize]{(1)} the obstacle size,
\hyperref[subsec:shortrange]{(2)} the short-range repulsion,
\hyperref[subsec:higherorder]{(3)} the different hydrodynamic moments,
and \hyperref[subsec:nearfield]{(4)} near-field flow.

\subsection{Effects of Obstacle Size}
\label{subsec:obstaclesize}

State diagrams for a representative selection of obstacle sizes $R$
and constant $r_\text{cut}=0$ are shown in \cref{fig:vor-R}.  At the
smallest obstacle, $R=R_\text{S}$, one observes that strong pushers
($\beta<0$) enter into forward orbits.  The critical value of $\beta$
below which the squirmer is captured is constant below
$y_0\approx 1.5R$ and decreases beyond this point.  Between the
orbiting and the scattering states lies a hovering state that also
extends to $y_0\approx 1.5R$.

As $R$ increases, one first observes that the $y_0$ required to
capture the squirmer decreases.  Simultaneously, the forward orbits
are mostly replaced with backward orbits, though a small region of
forward orbiting remains, and for $R\approx 2R_\text{S}$ a second
hovering region appears, between the backward orbits and the hovering
states (see inset in \cref{fig:vor-R}).  This region of forward orbits
quickly shrinks as $R$ increases and corresponds to a set of edge-case
trajectories, e.g., ones where the squirmer moves just slightly faster
than the criterion we picked to delimit orbiting from hovering.

Further increase of $R$ introduces a forward orbiting state for strong
pullers ($\beta>0$); the critical $\beta$ that separates these forward
orbits from scattering is independent of $y_0$ for $y_0<R$ and
decreases as $R$ increases.  Furthermore, at $y_0\approx R$, the
forward orbiting state extends significantly into the scattering
state's region.  This peninsula of red in \cref{fig:vor-R} appears
because the squirmer approaches the obstacle in such a way that no
hydrodynamic reorientation is required to swing into orbit.

Finally, one observes a second forward orbiting state that develops
for strong pushers.  This one is near $y_0\approx R$, unlike the other
states, which can be entered at $y_0\approx 0$.  At $R=200$, the state
diagram is already almost indistinguishable from the case for a flat
wall ($R\rightarrow\infty$).  In the latter case, $y_0$ becomes
meaningless and is replaced by $\varphi_0$, which is a well-defined
quantity even for finite radii, but cannot describe $y_0>R$.

\begin{figure}[tb]
\centering
\input{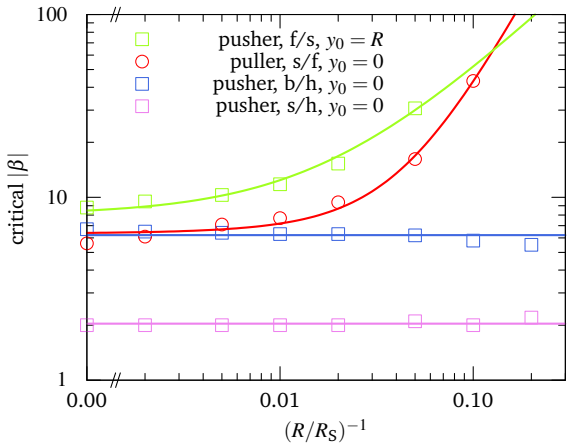}
\caption{From \cref{fig:vor-R} we can extract critical values of
  $\beta$ that mark the transition from one state to another.  Blue is
  the transition between backward orbiting (b) and hovering (h),
  violet is the transition between hovering and scattering (s), and
  red is the transition between hovering and forward orbiting (f), all
  near $y_0=0$.  Green is the transition from forward orbiting to
  scattering near $y_0=R$.  The error bars are comparable to the symbol size and result from the finite size
  of the regions produced by the Voronoi construction and also from
  the slightly diffuse transition regions (see the inset of
  \cref{fig:vor-R}).  }
\label{fig:points}
\end{figure}

In \cref{fig:points}, we have extracted the position of the critical
values of $\beta$ for the transitions observed in \cref{fig:vor-R}.
The transitions from scattering to hovering and from hovering to
backward orbiting are nearly independent of $\beta$.  The transition
between scattering and forward orbiting happens at
$\beta\propto R^{-2}$ for pullers at $y_0\approx 0$ and at
$\beta\propto -R^{-1}$ for pushers at $y_0\approx R$.  This scaling
disagrees with \citeauthor{spagnolie15a}'s
prediction\cite{spagnolie15a} for the pusher,
$\beta\propto -R^{-1/2}$, but as discussed in
\cref{subsec:higherorder}, the deviation is fully explained by a
modeling difference.

\subsection{Effects of Short-Range Repulsion}
\label{subsec:shortrange}

\begin{figure*}[htb]
\centering
\input{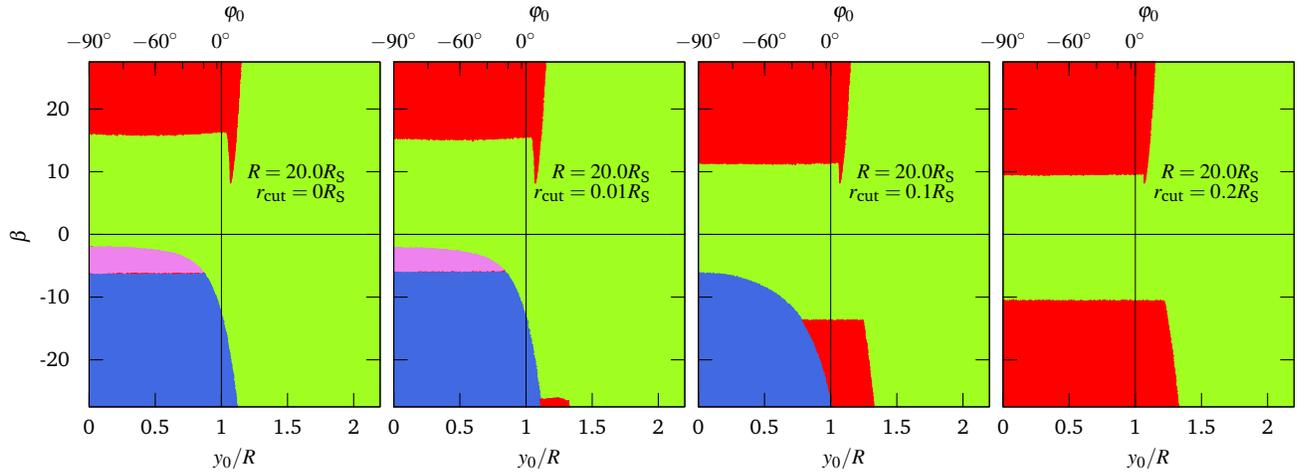}
\caption{ The influence of non-hydrodynamic interactions: State
  diagrams of the squirmer-obstacle interaction with color coding as
  in \cref{fig:vor-R}.  The dipolarity $\beta$ and initial position
  $y_0$ (or equivalently, initial orientation $\varphi_0$) are varied
  at constant $R=20R_\text{S}$.  From left to right we increase the
  sort-range repulsion, $r_\text{cut} = 0$, $0.01$, $0.1$, and
  $0.2 R_\text{S}$.  }
\label{fig:vor-gap}
\end{figure*}

Thus far, we have assumed $r_\text{cut}=0$, letting the squirmer and
obstacle touch.  However, realistic swimmers typically repel each
other and from obstacles at short distances\cite{lintuvuori16a}, e.g.,
due to electrostatics\cite{spagnolie12a}, phoretic
interactions\cite{popescu18a}, or near-field
hydrodynamics\cite{spagnolie12a}.  To study this effect, we pick
$R=20R_\text{S}$, where the state diagram contains all the features
seen at other obstacle radii.  We then construct iteratively refined
Voronoi diagrams for
$r_\text{cut}/R_\text{S}\in\left\{0,0.01,0.1,0.2\right\}$ in
\cref{fig:vor-gap}.

Making the step from $r_\text{cut}=0$ to $r_\text{cut}=0.01R_\text{S}$
introduces an additional length scale into the problem.  Despite the
small absolute magnitude of this $r_\text{cut}$, this leads to the
appearance of the transition from scattering to a forward orbit for
pushers near $y_0=R$.  While one cannot see this forward orbiting
state in the diagram for $r_\text{cut}=0$, it is expected at
$\beta\approx-40$ per \cref{fig:points} and visible in
\cref{fig:vor-R} for larger $R$.  Increasing $r_\text{cut}$ further
moves both transitions between scattering and forward orbits to
smaller $\left|\beta\right|$.  At $r_\text{cut}=0.1R_\text{S}$, the
hovering state has vanished completely.  At
$r_\text{cut}=0.2R_\text{S}$, the backward orbiting state has vanished
too and is replaced by forward orbits, which have now extended to
smaller $y_0$.

The disappearance of the hovering state at relatively moderate
short-range repulsion is again in line with our attribution of the
observations to the quadrupole term, which at small $\beta$ can only
dominate for the smallest gap sizes.
Even $r_\text{cut}=0.2R_\text{S}$ is sometimes a realistic model for short-range repulsion,
for example for chemical nanoswimmers with extended electric double layers\cite{lee14a,brown14a}.
This could explain why backward orbits are not encountered more commonly in experiment and theory.

\subsection{Effects of Higher-Order Hydrodynamic Modes}
\label{subsec:higherorder}

\begin{figure}[htb]
\centering
\input{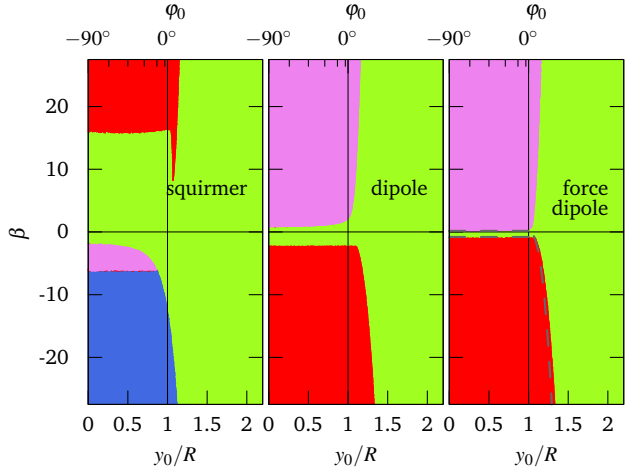} 
\caption{ The influence of higher-order hydrodynamic moments: State
  diagrams of the squirmer-obstacle interaction with color coding as
  in \cref{fig:vor-R}.  The obstacle radius is maintained at
  $R=20R_\text{S}$, while the dipolarity $\beta$ and initial position
  $y_0$ (or equivalently, initial orientation $\varphi_0$) are varied.
  From left to right we different hydrodynamic models are considered:
  squirmer, force and source dipole, and force dipole.  The gray
  dashed lines in the right panel indicate the position of the
  transition as predicted by Ref.~\citenum{spagnolie15a}.  }
\label{fig:vor-noquad}
\end{figure}

In the previous section, we have already observed that backward orbits
and hovering are very much dependent on near-field interactions.  This
even goes to the extent that backward orbits are completely suppressed
if squirmer and obstacle are kept sufficiently far apart.  The results
suggests that one of the higher hydrodynamic modes in \cref{eq:squirmerflow} causes
this behavior, since they dominate the flow only on short distances.
We perform two additional far-field calculations to quantify this
effect: (i) one that drops the source quadrupole from
\cref{eq:squirmer-decomposition,eq:reflected-squirmer-decomposition-1,eq:reflected-squirmer-decomposition-2}
but is otherwise identical to the method described in
\cref{sec:theory}, and (ii) one that furthermore drops the squirmer's
source dipole, leaving only the force dipole flow and moving the
squirmer directly via $v_0$.

Again, we pick $R=20R_\text{S}$ and $r_\text{cut}=0$ and obtain the
state diagram in \cref{fig:vor-noquad}.  One can see that the lack of
a quadrupole term replaces the backward orbits and hovering states
with forward orbits.  It also converts the forward orbits into a
hovering state down to much smaller $\beta$.  Further dropping the
source dipole allows for direct comparison with \citet{spagnolie15a},
who predict the position of the transition: They suggest that pushers
orbit for $\beta<-\sqrt{1024R_\text{S}/81R}$ as long as
$y_0<0.86\beta^{2/5}(R/R_\text{S})^{1/5}+R/R_\text{S}$, while they see
pullers hovering for $\beta>32R_\text{S}/9R$.  We reproduce these
predictions quite well as seen in \cref{fig:vor-noquad}.  The
remaining deviation is consistent with Ref.~\citenum{spagnolie15a},
where the critical $\left|\beta\right|$ is found to be slightly larger
than predicted.  Furthermore there is a slight difference in modeling,
namely that \citeauthor{spagnolie15a} do not include the swimmer's
finite $R_\text{S}$ in \cref{eq:faxen-v}.

\subsection{Accounting for the Near Field}
\label{subsec:nearfield}

\begin{figure}[htb]
\centering
\input{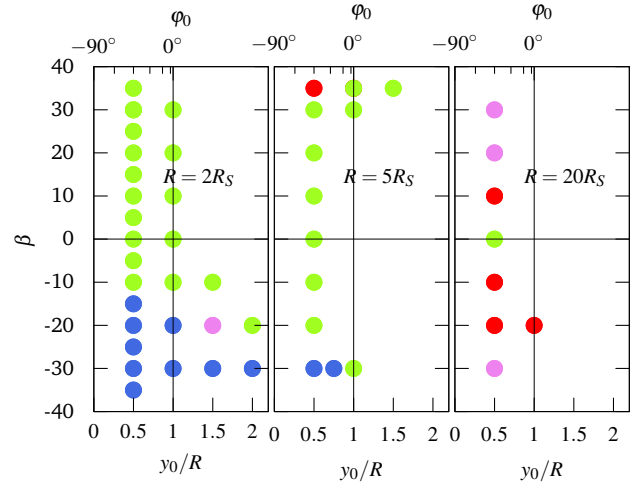} 
\caption{ The influence of near-field hydrodynamics: State diagrams
  obtained via the LB method indicate the interaction between the
  squirmer and obstacle as a function of dipolarity $\beta$ and
  initial position $y_0$ (or equivalently, initial orientation
  $\varphi_0$); the color coding is as in \cref{fig:vor-R}.  From left
  to right three values of $R = 2$, $5$, and $20 R_\text{S}$ are
  considered.  Compare to \cref{fig:vor-R} to see the differences
  caused by the near field.  }
\label{fig:vor-LB}
\end{figure}

Finally, we come to the results obtained using our LB calculations,
see \cref{subsec:LB}.  These are computationally much more involved
and we were therefore only able to sample a few points to verify the
general trends of our far-field prediction.  We restrict ourselves to
$R/R_\text{S} \in\left\{2,5,20\right\}$ and a few values of $\beta$
and $y_0$.  The state diagram in \cref{fig:vor-LB} shows that
scattering is generally more prevalent than in the previous
far-field-only calculations.  Backward orbits are still observed, but
their critical dipolarity increases more rapidly with decreasing
curvature.  Forward orbits, which in the far field are primarily
predicted for pullers, are now found in the case of sufficiently
strong pushers when the obstacle size is large enough.  Hovering
states are possible for both pushers and pullers of sufficient
strength, whenever the curvature is low.

These observations confirm that all four states found in the far-field
model are indeed allowed by the near-field flow.  Most notably,
backward orbits of strong pushers appear in both far-field and LB
models.  However, agreement with the far-field model and with
lubrication theory is only obtained to a certain degree, as expected.
The observed rapid decrease of $\beta$ with $R$ for backward orbiting
places the backward sliding state ($R \rightarrow \infty$) outside the
capabilities of our LB calculations.

\section{Conclusion}
\label{sec:conclusion}

We have employed three hydrodynamic methods to investigate the
behavior of a squirmer near a spherical obstacle or a flat wall.  Our
primary results are derived using a simple far-field approximation,
which lends itself to an efficient exploration of parameter
space.  Depending on the squirmer dipolarity, incidence angle,
obstacle curvature, and short-range repulsion, this revealed four classes
of trajectories: scattering, forward orbits, backward orbits, and
hovering.  Three of these trajectory classes have been previously
reported, but the backward orbits constitute a new class that
appears only for strong pushers.

Using the far-field approximation allowed us to construct state
diagrams that cover the entire parameter space.  We obtained all four
classes for reasonable dipolarity parameters whenever the squirmer
size is less than roughly half the size of the obstacle.  Comparison to calculations that exclude
the squirmer's quadrupole term reconcile our results with those of
\citet{spagnolie15a}.  This clearly attributes hovering and backward
orbiting to the quadrupole.
Thus, accounting for the finite size of the swimmer in the hydrodynamic
multipole expansion introduces a richer behavior.

We also computed trajectories for several parameter
sets using the LB method to investigate whether the
reported far-field behaviors persist even when taking into account
near-field details.  While the exact positions of the transition
between classes are altered, the qualitative behavior stays the same.
Most importantly, we reproduce the predicted backward orbiting for the
strong pushers in these calculations, showing that this effect is not
an artifact of our approximation.

Our results indicate a mechanism of mobility reversal with respect to
the bulk that is exclusively due to the hydrodynamic interaction of a
swimmer with a surface.  However, biological or artificial swimmers
may additionally interact chemically or electrostatically with
surfaces.  Simple mappings of chemical swimmers onto a
squirmer\cite{michelin14a,ibrahim16a,popescu18a} are known not to
qualitatively capture their behavior at small
separation\cite{ibrahim16a,popescu18a}.  Thus, such effects need to be
accounted for in unison with the hydrodynamics.  The present theory
and LB calculations provide a stepping stone toward analysis of
(electro)chemical contributions to the obiting of artifical swimmers.

\section*{Conflicts of interest}
There are no conflicts to declare.

\section*{Acknowledgements}
We acknowledge the Deutsche Forschungsgemeinschaft (DFG) for funding
through the SPP 1726 ``Microswimmers: from single particle motion to
collective behavior'' (HO1108/24-2).  JdG further acknowledges funding
through the NWO START-UP grant (740.018.013).  Computational resources
were provided by the state of Baden-W\"urttemberg through bwHPC and by
the DFG through grant INST 35/1134-1 FUGG.  We are grateful to
Alexander Chamolly and Will Uspal for useful discussions.

\section*{Research Data}
The numerical code and analysis scripts used to obtain the data presented in this publication
is available at \url{https://doi.org/10.24416/UU01-BMGD4E}, along with a representative subset of the data.


\raggedright
\bibliographystyle{jabbrv_rsc}
\bibliography{bibtex/icp,squirmer}


\appendix

\section{Hydrodynamic Reflection at a Sphere}
\label{app:reflection}

\justify
In this appendix we reproduce the reflection flow field of a Stokeslet near a spherical obstacle with a no-slip boundary condition (\cref{eq:noslip}) located at the origin\cite{oseen27a,spagnolie15a}.
The Stokeslet flow (\cref{eq:stokeslet}) originates at $\vec{r}_\text{S}$ such that the image tensor is given by
\begin{widetext}
\begin{align}
\mathcal{R}\mathcal{M}(\vec{r},\vec{r}_\text{S}) =
-&\frac{R}{r_\text{S}\dr_*} \mathbb{1} -\frac{R^3}{r_\text{S}^3\dr_*^3} \drv_*\otimes\drv_* -\left(r^2-R^2\right)\Phi \nonumber \\
-&\frac{r_\text{S}^2-R^2}{r_\text{S}} \left( \frac{1}{R^3\dr_*} \vec{r}_*\otimes\vec{r}_* - \frac{R}{r_\text{S}^2\dr_*^3}\left( \vec{r}_*\otimes\drv_* + \drv_*\otimes\vec{r}_* \right) + \frac{2\vec{r}_*\cdot\drv_*}{R^3\dr_*^3} \vec{r}_*\otimes\vec{r}_* \right) ,
\intertext{with}
\Phi=
\phantom{+}&\frac{r_\text{S}^2-R^2}{2r_\text{S}^3}\left[-\frac{3}{R\dr_*^3}\drv_*\otimes\vec{r}_\text{S}
+\frac{R}{\dr_*^3}\mathbb{1} \right. -\frac{3R}{\dr_*^5}\drv_*\otimes\drv_*
-\frac{2}{R\dr_*^3}\vec{r}_*\otimes\vec{r}_\text{S} +\frac{6\vec{r}_*\cdot\drv_*}{R\dr_*^5}\drv_*\otimes\vec{r}_\text{S} \nonumber\\
+&\frac{3R}{\dr_*r_*+\vec{r}\cdot\vec{r}_*-r_*^2}
\left(\frac{1}{r_*^2\dr_*}\drv_*\otimes\vec{r}_* +\frac{1}{\dr_*^3}\drv_*\otimes\drv_*+\frac{\dr_*-r_*}{r_*\dr_*}\mathbb{1}
\right)  \nonumber\\
-&\frac{3R}{r_*^2\dr_*^2\left(\dr_*r_*+\vec{r}\cdot\vec{r}_*-r_*^2\right)^2}\left(\left(r_*\drv_* +\dr_*\vec{r}_*\right)\otimes\left(\dr_*\vec{r}_\text{S}-r_*^2\drv_*+\dr_*r_*\left(\vec{r} -2\vec{r}_*\right)\right)\right) \nonumber\\
-&\left.\frac{3R}{r_*^2r\left(rr_*+\vec{r}\cdot \vec{r}_*\right)}\left(\vec{r}\otimes\vec{r}_* + rr_*\mathbb{1}\right)
+\frac{3R}{r_*^2r\left(rr_*+\vec{r}\cdot \vec{r}_*\right)^2}\left(r_*\vec{r} +r\vec{r}_*\right)\otimes\left(r_*\vec{r}+r\vec{r}_*\right) \right] ,
\end{align}
\end{widetext}
where the variables $\vec{r}_*= (R^2 / r_\text{S}^2) \vec{r}_\text{S}$ and $\drv_*=\vec{r}-\vec{r}_*$ have been introduced.
For the limiting case of a flat wall ($R \rightarrow \infty$), the former has the geometric interpretation of the location of the image Stokeslet.

\end{document}